\documentclass{article}  

\usepackage{graphicx}

\usepackage{geometry}
\usepackage{graphicx}
\usepackage[round]{natbib}
\usepackage{color}
\usepackage{amssymb}
\usepackage{amsmath}
\usepackage{lineno}
\usepackage{txfonts}
\usepackage{amstext}
\usepackage{mathabx}
\usepackage{multirow}
\usepackage{ulem}
\usepackage{lscape}
\usepackage{txfonts}
\usepackage{subfigure}
\usepackage{float}
\usepackage{chngpage}

\usepackage{hyperref}

\begin{document} 


\Large
\begin{center}\textbf{Impact of the transport of magnetospheric electrons on the composition of the Triton atmosphere}\end{center}\normalsize

\large\noindent B. Benne$^{1,2}$, B. Benmahi$^{3}$, M. Dobrijevic$^{1}$, T. Cavali\'e$^{1,4}$, J-C. Loison$^{5}$, K. M. Hickson$^{5}$, M. Barthélémy$^{6}$, J. Lilensten$^{6}$\normalsize\\
\vspace{0.2cm}

\noindent$^1$Laboratoire d'Astrophysique de Bordeaux, Univ. Bordeaux, CNRS, B18N, all\'ee Geoffroy Saint-Hilaire, 33615 Pessac, France \\
$^2$The University of Edinburgh, School of Geosciences, Edinburgh, United Kingdom (email: benjamin.benne@ed.ac.uk)\\ 
$^3$Laboratoire de Physique Atmosphérique et Planétaire, B5C Quatier Agora, Allée du Six Août 19c, 4000 Liège, Belgium\\
$^4$LESIA, Observatoire de Paris, Université PSL, CNRS, Sorbonne Université, Université Paris Cité, 5 place Jules Janssen, 92195 Meudon, France\\
$^5$Institut des Sciences Moléculaires, CNRS, Univ. Bordeaux, 351 Cours de la Libération, 33400 Talence, France\\
$^6$Univ. Grenoble Alpes, CNRS, IPAG, 38000 Grenoble, France\\

\vspace{0.2cm}
\noindent\textbf{Received:} 19 April 2023\\
\noindent\textbf{Accepted:} 23 February 2024\\
\vspace{0.2cm}

\noindent\textbf{DOI:} https://doi.org/10.1051/0004-6361/202346699 \\
\vspace{0.5cm}

\section*{Abstract}
Due to its inclined orbit and the complex geometry of the magnetic field of Neptune, Triton experiences a highly variable magnetic environment. As precipitation of magnetospheric electrons is thought to have a large impact on the Triton atmosphere, a better understanding of the interaction between its atmosphere and the magnetosphere of Neptune is important.
We aim to couple a model of the Triton atmosphere with an electron transport model to compute the impact of a varying electron precipitation on the atmospheric composition.
We coupled a recent photochemical model of the Triton atmosphere with the electron transport model TRANSPlanets. The inputs of this code were determined from Voyager 2 observations and previous studies. The main inputs were the electron precipitation flux, the orbital scaling factor, and the magnetic field strength. The electron-impact ionization and electron-impact dissociation rates computed by TRANSPlanets were then used in the photochemical model.
We also analyzed the model uncertainties.
The coupling of the two models enabled us to find an electron density profile, as well as N$_2$ and N number densities, that are consistent with the Voyager 2 observations. 
We found that photoionization and electron-impact ionization are of the same order, in contrast to the results of previous photochemical models. However, we emphasize that this result depends on the hypotheses we used to determine the input variables of TRANSPlanets. Our model would greatly benefit from new measurements of the magnetic environment of Triton, as well as of the electron fluxes in the Neptune magnetosphere.


\section{Introduction}
\label{Intro}

Since the flyby of Triton by Voyager 2 in 1989, we have known that the Triton ionosphere is surprisingly more dense than that of Titan. As Triton is three times farther from the Sun than Titan, it is difficult to explain this density. Thus, an additional ionization source seems to be required.
This supplementary source was quickly thought to be electrons precipitating from the Neptune magnetosphere \citep{majeed_ionosphere_1990,strobel_magnetospheric_1990}, as energetic particles were observed by Voyager 2 in the Neptune magnetosphere \citep{krimigis_hot_1989}. In addition, these precipitating electrons could bring the additional energy needed to explain the observed thermospheric temperatures \citep{stevens_thermal_1992,krasnopolsky_temperature_1993}. This precipitation is considered in the majority of Triton atmospheric models \citep{summers_tritons_1991,krasnopolsky_temperature_1993,krasnopolsky_photochemistry_1995,strobel_tritons_1995,strobel_comparative_2017,benne_photochemical_2022,krasnopolsky_tritons_2023}, with the exception of \citep{lyons_solar_1992}. The latter model did not consider electron precipitation, but was able to reproduce the electron profile by using an unrealistically high reaction rate for the charge exchange between N$_2^+$ and C. 

The magnetic environment of Triton is highly variable due to the combination of its inclined orbit and the complex configuration of the magnetic field of Neptune. 
As Voyager 2 only passed through the Neptune magnetosphere once, we lack crucial data to precisely describe this environment. \citet{strobel_magnetospheric_1990} used measurements performed by the Plasma Science System (PLS) and by the Low-Energy Charged Particles (LECP) instruments in the vincinity of Triton to determine the electron precipitation flux and compute the electron-impact ionization profile and the power deposited in the Triton atmosphere. To take into account the variability of the environment along the Triton orbit, they also used an orbital scaling factor of 0.25, which reduced the total amount of power deposited in the atmosphere. \citet{summers_tritons_1991} recommended shifting the ionization profile of \citet{strobel_magnetospheric_1990} upward by two scale heights ($\approx$100\,km) so that the maximum electron production rate would correspond to the electron density peak measured by Voyager 2. This modification was then used in the models of \citet{stevens_thermal_1992}, \citet{krasnopolsky_temperature_1993}, \citet{krasnopolsky_photochemistry_1995}, \citet{strobel_tritons_1995}, and \citet{benne_photochemical_2022}. This shift of the ionization profile is equivalent to considering that high-energy electrons ($E>50$\,keV) do not precipitate in the Triton atmosphere \citep{stevens_thermal_1992}. However, this hypothesis was questioned by \citet{sittler_tritons_1996}, whose results suggest that high-energy electrons are the most likely to precipitate in the Triton atmosphere. 
In their own model, \citet{krasnopolsky_temperature_1993} considered the orbital scaling factor as a free parameter. They chose to use a factor of 0.162 in their nominal model, with which they multiplied the ionization profile from \citet{strobel_magnetospheric_1990} to fit the electron density observations from Voyager 2. Therefore, ad hoc modifications were made in previous models of the Triton atmosphere to model the electron precipitation and find electron density profiles consistent with the Voyager 2 observations.

\citet{benne_photochemical_2022} found that electron-impact ionization was the main source of ionization in the Triton atmosphere, with a ratio of the photoionization to electron-impact ionization of 3/8. However, their maximum electron density was 2.5 to 5 times higher than the measured peaks presented in \citet{tyler_voyager_1989}, even when chemical uncertainties were considered. Thus, it appeared necessary to model the precipitation of magnetospheric electrons better in order to improve these results. This is the aim of the present work: We couple the photochemical model of \citet{benne_photochemical_2022} with the electron transport model TRANSPlanets. This model is derived from the family of the TRANS* models that were used to compute electron transport in various planetary atmospheres \citep{gronoff_ionization_2009}. In this paper, we adapt it to the Triton conditions. Its main input parameters (electron precipitation flux, magnetic field strength, and orbital scaling factor) are computed from a model of the Neptune-Triton system and from previous studies \citep{strobel_magnetospheric_1990,sittler_tritons_1996}. TRANSPlanets then computes the electron-impact ionization and dissociation rates that are used in the photochemical model of the Triton atmosphere.
We present how we treated the highly variable magnetic environment of Triton in order to determine the inputs of the electron transport model in Sect. \ref{Trit_envrnmt}. We detail TRANSPlanets, its inputs, and how we coupled it to our photochemical model in Sect. \ref{sec_TRANSmodel}. We show the secondary ionization profiles in Sect. \ref{sec_e_prod} and our results, both nominal and considering chemical uncertainties, in Sect. \ref{sec_NomRes}, and we  also discuss them. We present a sensitivity study of the electron temperature in Sect. \ref{sect_sens_elec_temp} and finally give our conclusions in Sect. \ref{conclu}. 

\section{Variable magnetic environment of Triton}
\label{Trit_envrnmt}

The Neptune magnetic field is very complex, as it cannot be represented at close distances without considering quadrupole and octupole terms \citep{ness_magnetic_1989,connerney_magnetic_1991,ness_neptunes_1995}. When it is approximated with a dipole (this approximation is valid from 4 to 15 Neptunian radii ($R_{\text{N}}$) \citep{ness_neptunes_1995}, with $R_{\text{N}} = 24\,765$\,km), its center is offset from the center of the planet, and the dipole axis is not aligned with the rotation axis of the planet. Thus, the simple models of the magnetic field of Neptune are offset tilted dipole (OTD) models. The first model of this type was presented in \citet{ness_magnetic_1989} and is referred to as OTD1 throughout this paper. An updated version of this model was provided in \citet{ness_neptunes_1995} and is referred to as OTD2 in the following. Finally, \citet{connerney_magnetic_1991} reported a more complex model of the magnetic field of Neptune with quadrupole and octupole moments. Because these terms are only predominant at close distances from Neptune, however, they also give a dipolar approximation of their model for computations at larger distances. We refer to this model as OTD-O8. The parameters of the three models are given in Table \ref{compar_OTD}, which is taken from \citet{ness_neptunes_1995}. 

\begin{table}[!h]
  \centering
     \begin{tabular}{@{}ccccccc@{}}
        \hline 
        Model & \begin{tabular}[c]{@{}c@{}}Dipole moment\\ {(}T.$R_{\text{N}}^3${)}\end{tabular} & \begin{tabular}[c]{@{}c@{}}$x_c$\\ {(}$R_{\text{N}}${)}\end{tabular} & \begin{tabular}[c]{@{}c@{}}$y_c$\\ {(}$R_{\text{N}}${)}\end{tabular} & \begin{tabular}[c]{@{}c@{}}$z_c$\\ {(}$R_{\text{N}}${)}\end{tabular} & \begin{tabular}[c]{@{}c@{}}Tilt angle\\ {(}°{)}\end{tabular} & \begin{tabular}[c]{@{}c@{}}Direction\\ {(}° W{)}\end{tabular} \\ \hline 
        OTD1$^1$ & 1.33\,10$^{-5}$ & 0.17 & 0.46 & -0.24 & 46.8 & 79.5 \\
        OTD2$^2$ & 1.30\,10$^{-5}$ & 0.19 & 0.48 & -0.19 & 45.2 & 76.5 \\
        OTD-O8$^3$ & 1.42\,10$^{-5}$ & 0.05 & 0.48 & 0.00 & 46.9 & 72.0 \\ \hline 
        \end{tabular}
     \caption{ Comparison between the different OTD models used to approximate the magnetic field of Neptune. \\ \textbf{Notes:} $^1$ \citet{ness_magnetic_1989} ; $^2$ \citet{ness_neptunes_1995} ; $^3$ \citet{connerney_magnetic_1991}.\\ ($x_c,y_c,z_c$) are the coordinates of the center of the dipole, using the coordinate system defined by the Voyager Project Steering Group \citep{connerney_magnetic_1991}. This coordinate system differs from that of the International Astronomical Union \citep[see for example][]{archinal_report_2018}. The tilt angle corresponds to the inclination of the magnetic dipole axis with respect to the Neptune rotation axis, and the direction indicates toward which longitude the dipole axis is inclined. The west longitude is defined as in \citet{connerney_magnetic_1991}.}
     \label{compar_OTD}
\end{table}

The offset-tilted nature of this magnetic dipole, coupled to the inclination of the Triton orbit and to the fast rotation of the planet (16.1\,h) in comparison to the orbital period of Triton (5.88 Earth days), causes the satellite distance to the center of the dipole, as well as its magnetic latitude, to vary strongly.
This can be observed when computing the $L$-shell parameter of Triton. This parameter corresponds to the distance at which a given field line of a dipolar field crosses the magnetic equator \citep{ness_magnetic_1989}. In our case, this corresponds to the $L$-shell of the line going through the position of Triton. For a dipole field, this parameter is obtained for the following formula:

\begin{equation}
   L(\lambda, r) = \frac{r}{\cos^2{(\lambda)}},
\end{equation}
\noindent where $r$ is the distance between Triton and the center of the dipole in $R_{\text{N}}$, and $\lambda$ is the magnetic latitude.

\begin{figure*}[!h]
  \resizebox{\hsize}{!}
           {\includegraphics{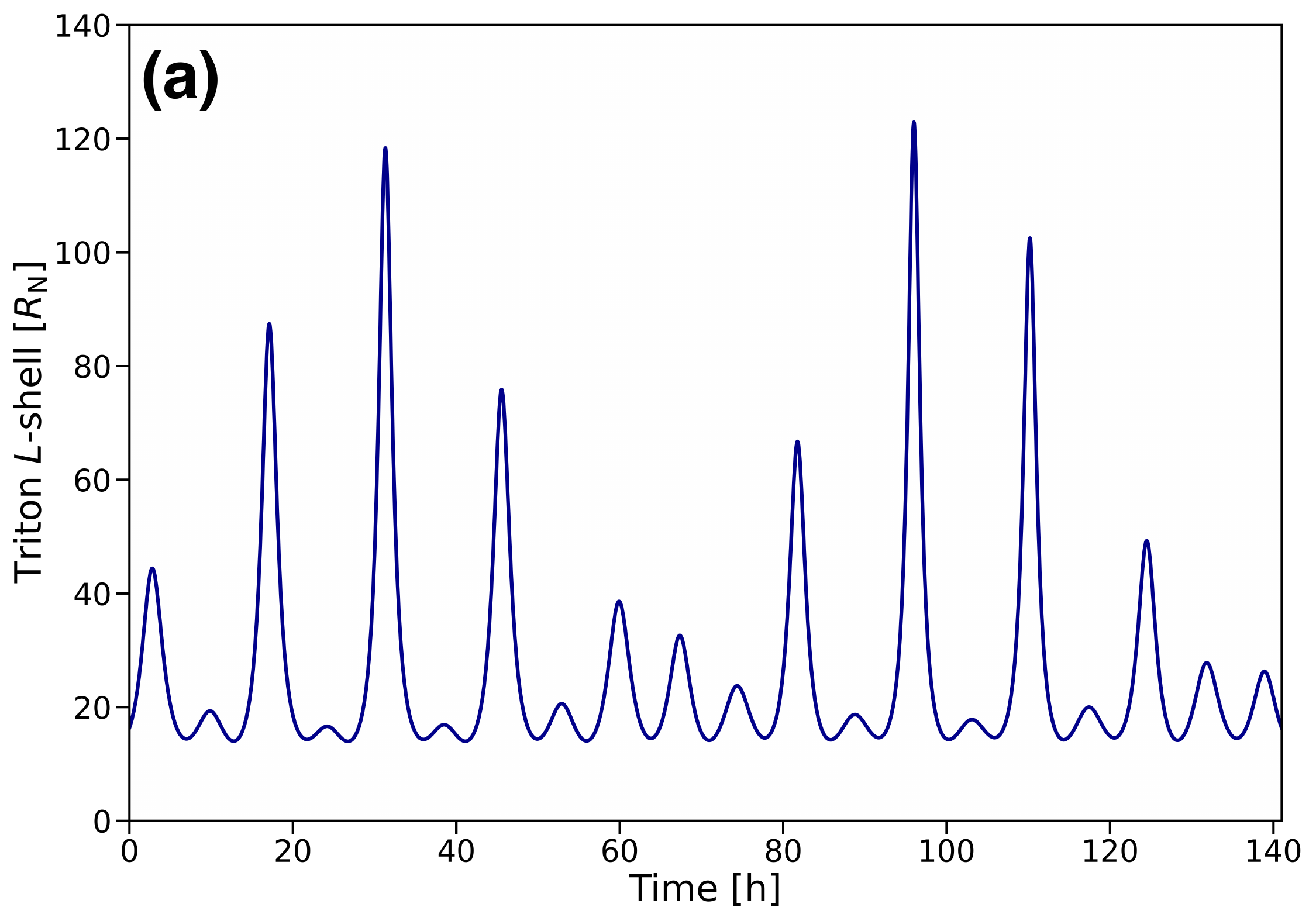}
           \includegraphics{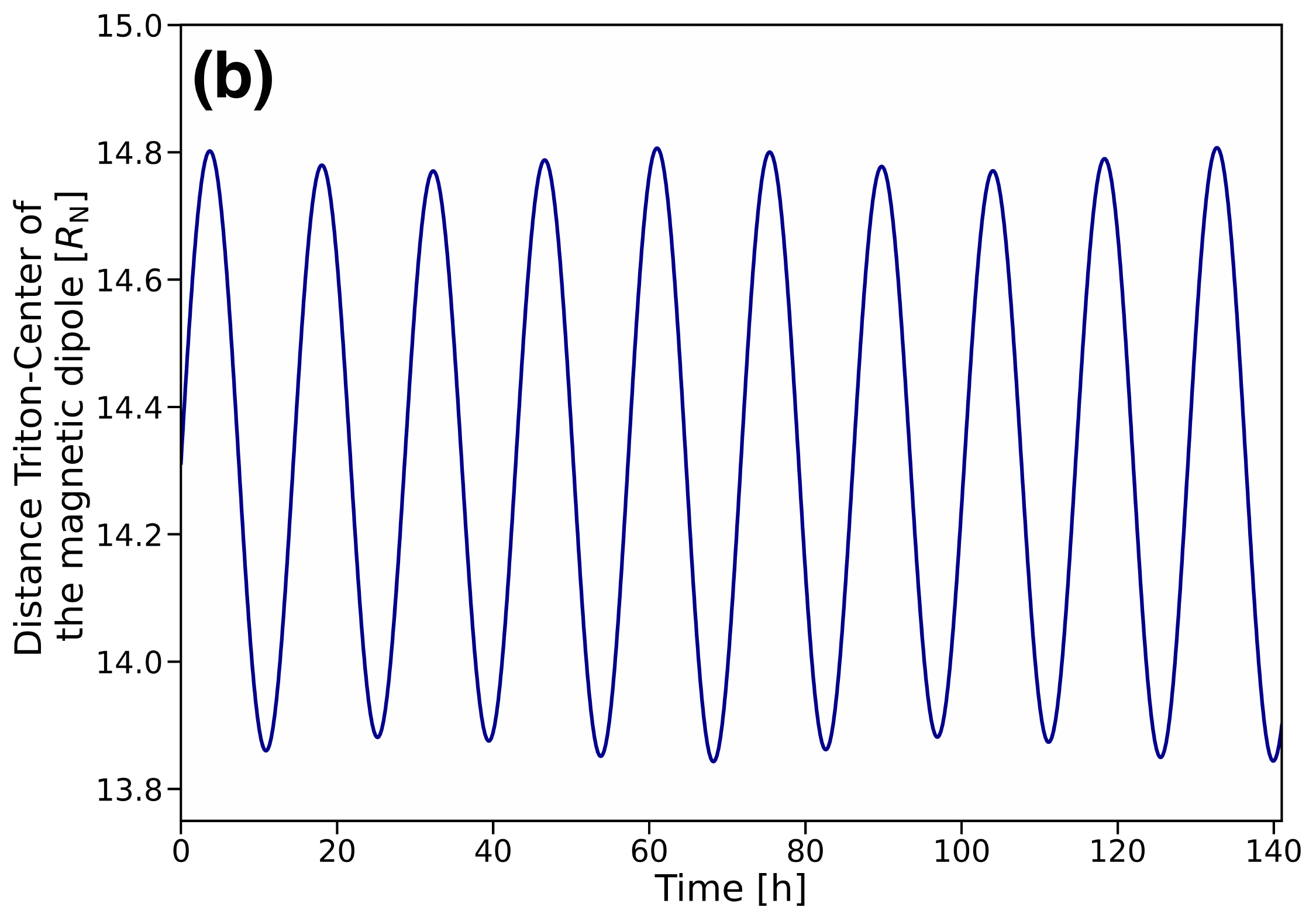}
           }
     \caption{Examples of the variation in some parameters used to describe the magnetic environment of Triton during one orbit around Neptune. These results are obtained with the OTD-O8 model from \citet{connerney_magnetic_1991}, starting from August 21, 1989. \textbf{(a)} Variation in the $L$-shell parameter of Triton. \textbf{(b)} Variation in the distance between Triton and the center of the dipole.
             }
        \label{d_TriMagnCtr+L_shell}
\end{figure*}

To compute the different parameters needed to describe the magnetic environment of Triton, we developed a Python code. This code allows us to derive the position of Triton with respect to Neptune and to the magnetic dipole, using the ephemeris from the JPL \textit{Horizons System}\footnote{\url{https://ssd.jpl.nasa.gov/horizons/app.html\#/}}. We found that $\lambda$ ranges from roughly -70$^{\circ}$ to +70$^{\circ}$, which causes $L$ to vary from 14 to more than 100 for every OTD model (128 for the OTD1 and OTD-O8 models, and 109 for the OTD2 model), as shown in panel (a) of Fig. \ref{d_TriMagnCtr+L_shell}. 
As the inner magnetosphere of Neptune seems to be delimited by the minimum $L$-shell of Triton \citep{mauk_energetic_1995} and as the electron number density is only assumed to be important near the magnetic equator \citep{strobel_magnetospheric_1990}, the electron precipitation into the Triton atmosphere may vary strongly along its orbit.

In panel (b) of Fig. \ref{d_TriMagnCtr+L_shell}, Triton remains within the distance interval in which the dipolar approximation is valid. Thus, knowing the magnetic latitude of Triton, the magnetic moment of the dipole, and the distance between its center and the satellite, we can compute the norm of the magnetic field as

\begin{equation}
   \left \lVert \overrightarrow{B} \right \rVert = \frac{M}{r^3}\times \sqrt{1 + 3\sin^2(\lambda)},
   \label{B_Field_eq}
\end{equation}
\noindent with $M$ as the magnetic moment in T.$R_{\text{N}}^3$ ($r$ is then expressed in $R_{\text{N}}$).
When we use the same time period as in Fig. \ref{d_TriMagnCtr+L_shell}, we obtain the magnetic field at Triton presented in panel (a) of Fig. \ref{B+histo_15,5}, which shows a factor of about two between extrema.

\begin{figure*}[!h]
   \resizebox{\hsize}{!}
            {\includegraphics{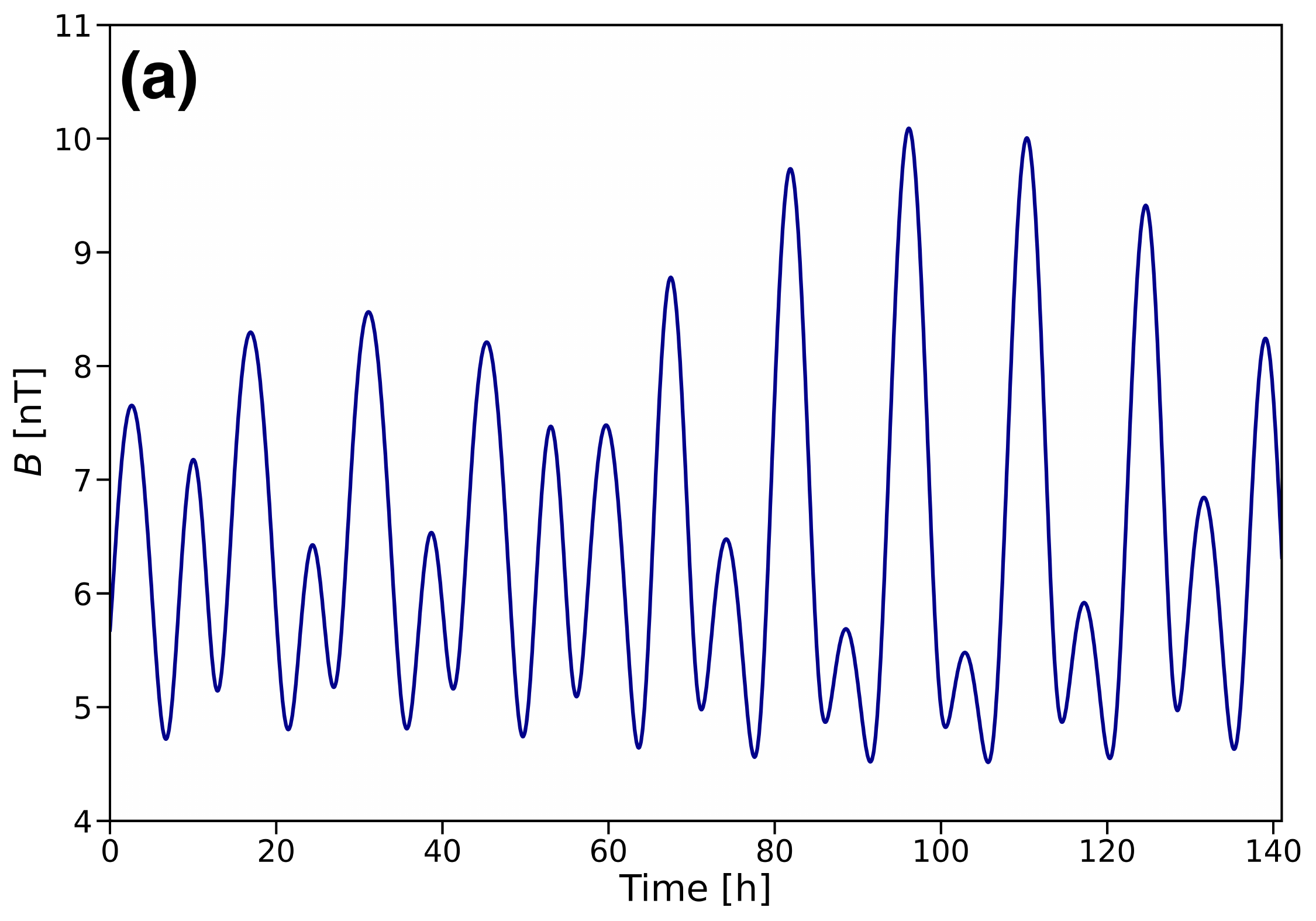}
            \includegraphics{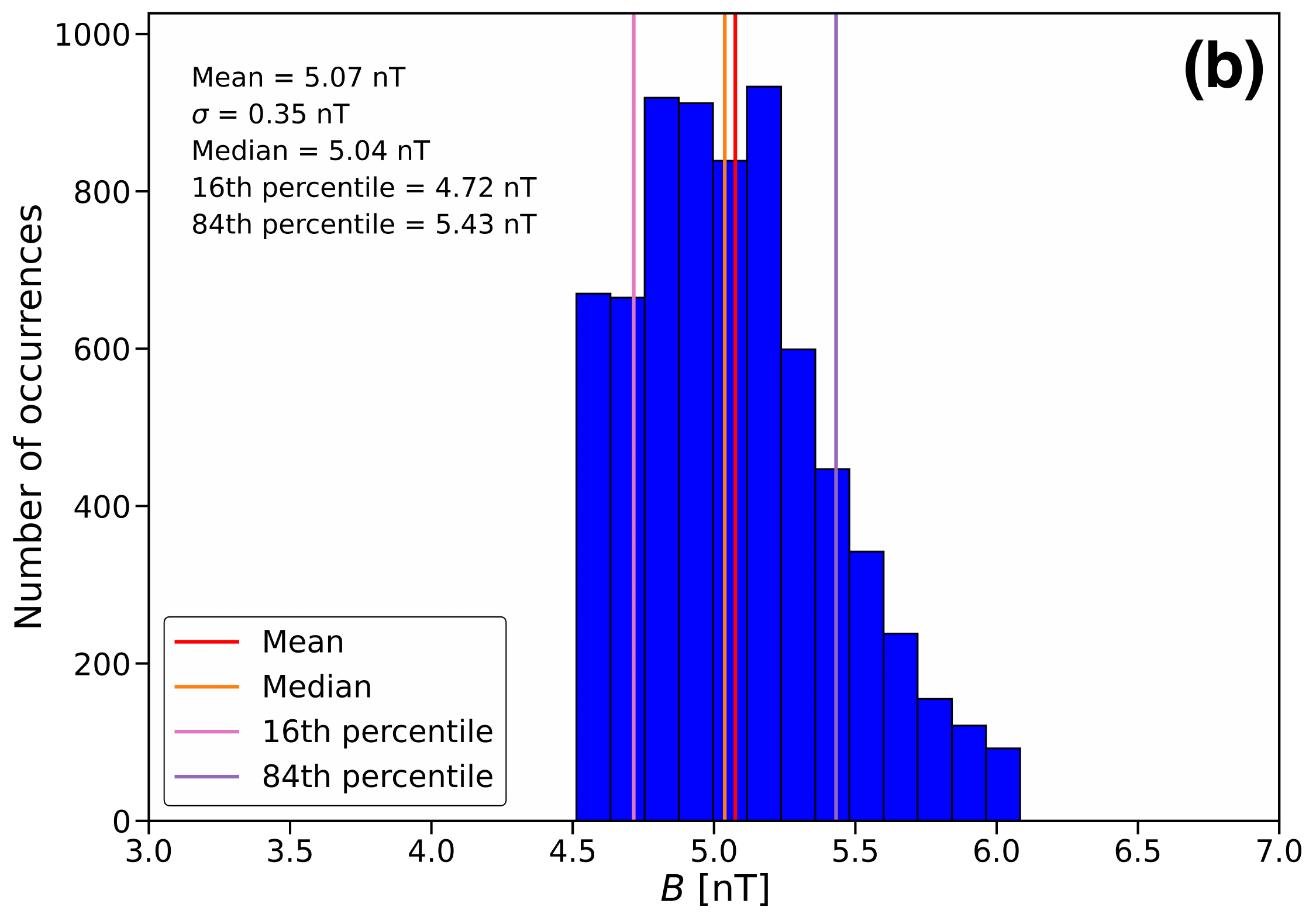}
            }
      \caption{Variation in the magnetic field at Triton. These results were obtained with the OTD-O8 model from \citet{connerney_magnetic_1991}. \textbf{(a)} Variation in the magnetic field norm over one orbit around Neptune, starting from August 21, 1989. \textbf{(b)} Histogram of the value of the norm of the magnetic field at Triton when the Triton $L$-shell is lower than 15.5. These results were computed over one period of variation in the magnetic field at Triton, i.e., 35 days, from August 21, 1989.
              }
         \label{B+histo_15,5}
\end{figure*}

\section{The TRANSPlanets model}
\label{sec_TRANSmodel}

TRANS is an electron transport model that was first developed to model the effects of electron precipitation into the Earth atmosphere \citep{lilensten_resolution_1989}. It was then adapted to a variety of Solar System bodies such as Venus \citep{gronoff_modelling_2007,gronoff_modelling_2008}, Mars \citep{witasse_prediction_2002,witasse_correction_2003,simon_dayglow_2009,nicholson_fast_2009}, Jupiter \citep{menager_modelisation_2011}, or Titan \citep{lilensten_fast_2005,lilensten_prediction_2005,gronoff_ionization_2009,gronoff_ionization_2009-1,gronoff_ionization_2011}. 

\begin{figure}[!h]
  \centering
  \includegraphics[width=0.65\textwidth]{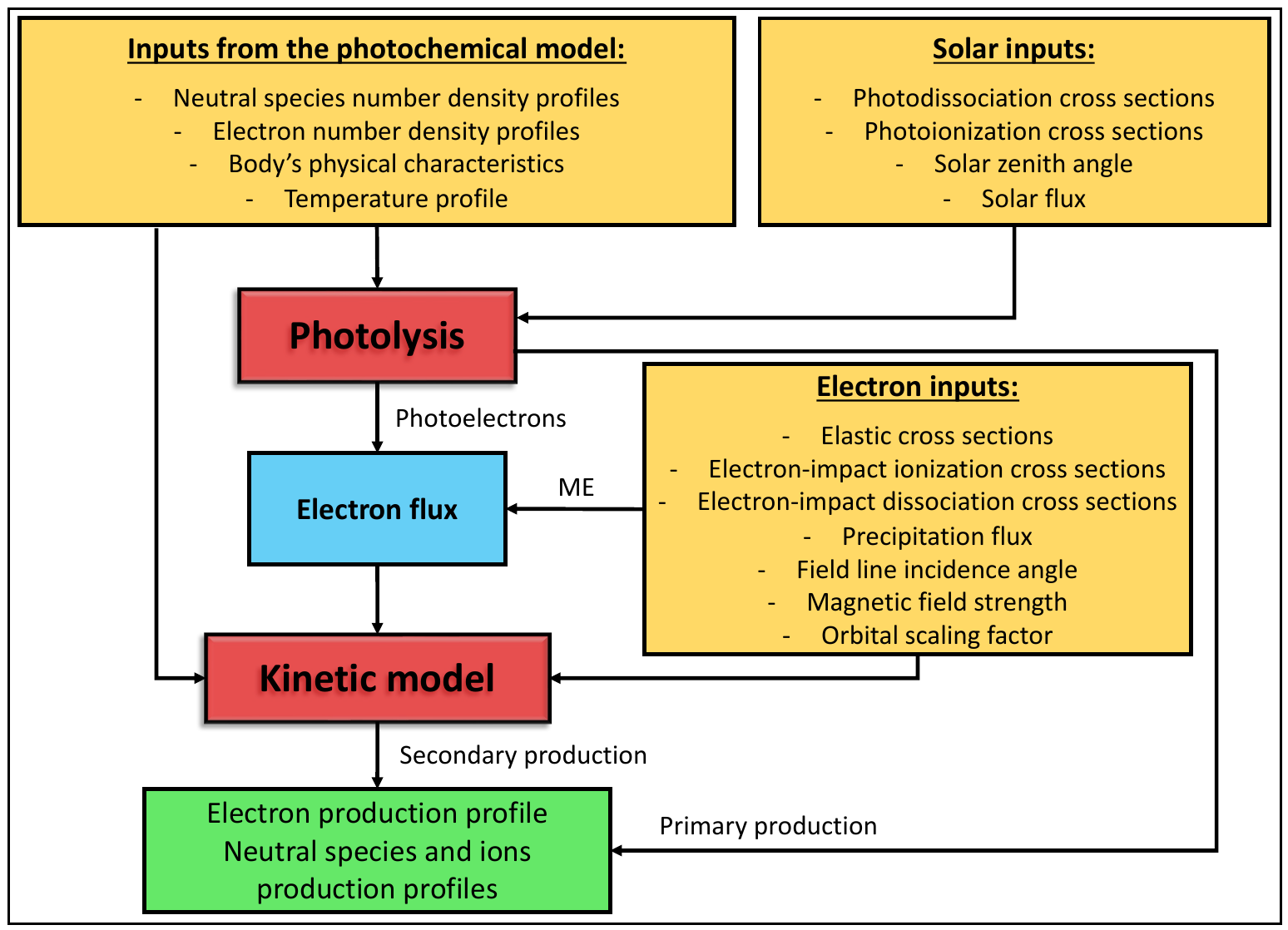}
  \caption{TRANSPlanets model workflow. The inputs are shown in the yellow boxes, the computational parts in the red boxes, the intermediate variables in the blue box, and the final outputs in the green box. ME stands for magnetospheric electrons.
   }
  \label{workflow_TRANS}
\end{figure}

The TRANS model computes the production of ions and electrons following the propagation of a flux of suprathermal electrons in the atmosphere. This flux is composed of photoelectrons produced by the photoionization of atmospheric species by solar extreme-UV (EUV) photons and of precipitating magnetospheric electrons. This primary ionization (i.e., photoionization) is also computed in TRANS. The code can be used with photoelectrons only, magnetospheric electrons only, or with both, as in this work. Then, the code computes the secondary ionization (ionization by the suprathermal electrons from primary ionization and electron precipitation) and dissociation through electron-impact reactions between this propagating flux and the atmospheric species. 
We describe the flow of operations of the code in Fig. \ref{workflow_TRANS}. The secondary production rates of the ions and electrons computed by TRANS are of interest in our case, as they can be used in our photochemical model. We consequently modified the original code used for Jupiter from \citet{menager_modelisation_2011} into a generic Fortran 90 version, which we call TRANSPlanets. This can be used for any planetary body and was adapted here to Triton.

In this section, we describe the fundamental equations used in the code, the inputs we used for Triton, and how we coupled it to our photochemical model. We point out that complete descriptions of the fundamental equations used in TRANS and TRANSPlanets have been presented by \citet{lilensten_resolution_1989}, \citet{simon_contribution_2006}, \citet{gronoff_etude_2009}, \citet{menager_modelisation_2011}, and \citet{benmahi_composition_2022}.

\subsection{Fundamental equations}
\label{subs_fundamental_eqs}

To compute the secondary ionization of the neutral species following the propagation of a suprathermal electron flux, TRANSPlanets solves the nonconservative Boltzmann equation, which describes the variation in the electron distribution function $f(\overrightarrow{r},\overrightarrow{v},t)$ in the phase space \citep{stamnes_inelastic_1983,karamcheti_kinetic_1968},

\begin{equation}
   \frac{\partial f}{\partial t} + \overrightarrow{v}\cdot \frac{\partial f}{\partial \overrightarrow{r}}+\frac{\partial }{\partial \overrightarrow{v}}\left ( \frac{f \overrightarrow{X}}{m_e} \right ) = S
   \label{boltz_eq_globale},
\end{equation}
where $\overrightarrow{r}$ is the position of the electrons, and $\overrightarrow{v}$ is their speed at time $t$. $\overrightarrow{X}$ represents the friction force between the suprathermal electrons and the thermalized electrons that are already part of the atmosphere, $m_e$ is the electron mass, and $S(\overrightarrow{r},\overrightarrow{v},t)$ is the source term describing the electron production. 
We express the friction force as in \citet{menager_modelisation_2011},

\begin{equation}
   \overrightarrow{X}=-n_eL(E)\frac{\overrightarrow{v}}{v}
   \label{Eq_friction_force},
\end{equation}

\noindent with

\begin{equation}
   L(E) = \frac{3.37\times 10^{-12}}{E^{0.94}n_e^{0.03}}\left ( \frac{E-E_{th}}{E-0.53E_{th}} \right )^{2.36}
   \label{Eq_friction_function},
\end{equation}
where $E$ is the energy of the suprathermal electron, $n_e$ is the number density of the thermal electrons, and $E_{th}$ is their energy. All the energies are expressed in eV. $L(E)$ is the friction function expressed in eV.cm$^2$ as given in \citet{swartz_analytic_1971} and recommended by \citet{stamnes_heating_1983}. 

Eq. \ref{boltz_eq_globale} can be expressed depending on the suprathermal electron flux $\Phi$ instead of on their distribution function by making the following variable change:

\begin{equation}
   \Phi(\overrightarrow{r},E,\overrightarrow{u},t)=\frac{v^2}{m_e}f(\overrightarrow{r},\overrightarrow{v},t)
   \label{eq_chgt_variable_boltz},
\end{equation}
with $E$ the kinetic energy of the electrons and $\overrightarrow{u}$ their direction of propagation, such that $\overrightarrow{u}=\frac{\overrightarrow{v}}{v}$, giving

\begin{equation}
   \frac{1}{v}\frac{\partial \Phi}{\partial t}+\frac{\overrightarrow{v}}{v}\cdot\frac{\partial \Phi}{\partial \overrightarrow{r}} - n_e \frac{\partial \left [ L(E) \Phi \right ]}{\partial E} = \frac{1}{v}\hat{S}
\end{equation}
with $\hat{S}(\overrightarrow{r},E,\overrightarrow{u},t)=\frac{v^2}{m_e}S(\overrightarrow{r},\overrightarrow{v},t)$. 

Considering steady state and a plane-parallel atmosphere, we obtain the equation describing the propagation of a flux $\Phi$ of suprathermal electrons along a magnetic field line projected on the vertical axis,

\begin{equation}
   \begin{split}
      \mu \frac{\partial \Phi(\tau,\mu,E)}{\partial \tau(z,E)} = - \Phi(\tau,\mu,E)+\frac{n_e(z)}{\sum_k n_k(z)\sigma_k^{tot}(E)}\frac{\partial \left [ L(E) \Phi(\tau,\mu,E) \right ]}{\partial E} \\
      + D(z,\mu,E) + P(z,\mu,E),
   \end{split}
   \label{Eq_boltz_Z}
\end{equation}
where $\mu$ is the cosine of the pitch angle, $\tau$ is the collision depth, which is similar to an optical depth in radiative transfer, $z$ is the altitude of the considered level, $n_k(z)$ is the number density of species $k$ at this level, and $\sigma_k^{tot}(E)$ is the total electron-impact cross section of species $k$ at energy $E$. The collision depth is computed as

\begin{equation}
   \tau(z,E) = \int_{z}^{z_{max}} d\tau ~~~~~~~ \text{with}~~ d\tau= \sum_k n_k(z)\sigma_k^{tot}(E)dz. 
\end{equation}
\noindent The suprathermal electrons center of mass propagates along the magnetic field line. The projection on the vertical axis is made by simply dividing by the cosine of the magnetic dip angle. This is incorrect for large angles (typically larger than 45$^{\circ}$) because the cosine tends toward zero. Rigorously, the most accurate results are obtained when the local magnetic field line coincides with the vertical dimension.

The source term is decomposed into two terms $P$ and $D$. $P$ is the electron production term, corresponding to the sum of the electron production from photoionization and the precipitation flux. $D$ is the diffusion term, resulting from the diffusion of suprathermal electrons to lower energies through collisions with atmospheric species. 
The code is multistream, meaning that $\mu$ is discretized on a certain number of angles chosen by the user. This allows computing the different output fluxes on various angles and not just the downward and upward fluxes. In our case, we chose to run our calculations over 16 angles over 360$^{\circ}$ because this discretization provided the best compromise in terms of accuracy and computational time.

\subsection{TRANSPlanets inputs}

As shown in Eq. \ref{Eq_boltz_Z}, TRANSPlanets needs some additional data to solve the Boltzmann equation. It needs the electron-impact ionization and dissociation cross sections for the considered atmospheric species, an input flux, an energy range, an energy grid, an altitude grid, the number density profiles of the considered species and of the thermal electrons, the temperature profiles for these species, and a solar flux with the associated photolysis cross sections. In the following sections, we first describe the inputs that are not related to the photochemical model. These inputs are detailed in Sect. \ref{couple_TRANS_PCM}. 

\subsubsection{Electron-impact cross sections and energy grid}

In our model, the electron-impact cross sections come from different sources, depending on the species. The elastic and inelastic (electron-impact ionization and dissociation) cross sections of N$_2$ were taken from the Atomic and Molecular Cross section for Ionization and Aurora Database (AtMoCiad) \citep{gronoff_drguiguiatmociad_2021}. Those of CO were taken from \citet{itikawa_cross_2015}. For C, we took the elastic cross sections from the NIST database, while the electron-impact ionization cross section was taken from \citet{jonauskas_electron-impact_2018} and the electron-impact excitation cross sections from \citet{suno_cross_2006}.
For the N($^4$S) and N($^2$D) cross sections, we used inelastic cross sections from \citet{jonauskas_electron-impact_2022} and \citet{kato_electron_1994}. The elastic cross sections below 50\,eV were taken from \citet{wang_b_2014}, and they were completed above this energy with data from the NIST database.
As the maximum energy used in the elastic cross section of N$_2$ in the AtMoCiad database is 10$^5$\,eV, we used this energy as the upper limit of our energy range. The lower boundary was set to 1\,eV to take into account rotational and vibrational excitation of N$_2$. The energy grid consisted of 600 points spaced logarithmically between these two energies, such that $E(i) = \exp\left [E(i-1) + \frac{\ln(E_{max})-\ln(E_{min})}{600}\right ]$. 

\subsubsection{Variable parameters}
\label{subs_variable_params}

The $L$-shell of Triton varies strongly with time, as shown in Fig. \ref{d_TriMagnCtr+L_shell}. \citet{krimigis_hot_1989} have shown that the electron flux decreases strongly outside of the minimum $L$-shell of Triton, this value presumably defining the limit of the inner magnetosphere of Neptune \citep{mauk_energetic_1995}. This led \citet{strobel_magnetospheric_1990} to consider that Triton experiences constant electron precipitation if $L(\text{Triton})<15.5$, with no precipitation for higher $L$ values. 
With this assumption, they computed an average precipitation flux of 25\% of the flux at the minimum $L$-shell of Triton. Thus, their orbital scaling factor was $m_{orb}=0.25$.

In their model, \citet{krasnopolsky_temperature_1993} considered $m_{orb}$ as a free parameter. Their best value to match the electron density profiles from Voyager 2 was $m_{orb}=0.162$. This value was used in their nominal model and then in \citet{krasnopolsky_photochemistry_1995}.
\newline 

With our modeling of the Neptune-Triton system, we are able to compute the value of the Triton $L$-shell at any moment. We could then have used a relation between the value of $L$ and the magnetospheric electron flux, but no such relation exists due to the scarcity of the Voyager 2 electron flux measurements. Thus, we decided to keep the criterion used in \citet{strobel_magnetospheric_1990} and considered that electron precipitation only occurs when the $L$-shell of Triton is lower than 15.5. Therefore, the orbital scaling factor corresponds to the ratio of the time when $L$ is lower than 15.5 to the time interval considered. 
We consistently find for all three magnetospheric models presented in Table \ref{compar_OTD} that $L$<15.5 for 27 to 28.5\% of the time. Therefore, we used $m_{orb}=0.27$.
\newline

The value of the magnetic field strength $B$ at Triton varies significantly with time, as shown in panel (a) of Fig. \ref{B+histo_15,5}. When we plot the histogram of the value of $\left \lVert \overrightarrow{B} \right \rVert$ when $L<15.5$, we obtain panel (b) of Fig. \ref{B+histo_15,5} (with the OTD-O8 model). The mean value of the magnetic field norm is (5.07$\pm$0.35)\,nT. This value is consistent with Fig. 3 of \citet{ness_magnetic_1989} and with the value found with Eq. \ref{B_Field_eq} using the values of $r$ and $\lambda$ from Fig. 7 of \citet{mauk_energetic_1995}.  This value is lower than the value used in \citet{strobel_magnetospheric_1990} and \citet{sittler_tritons_1996}, which is 8\,nT. 
This difference impacts the power deposited in the atmosphere by curvature drift, as shown in Sect. \ref{subs_prec_flux}.
When we assume that the electrons deposit their energy by curvature drift (as in \citealt{strobel_magnetospheric_1990}), using 5\,nT instead of 8\,nT leads to a higher deposited energy (see Eqs. \eqref{power_deposited} and \eqref{drift_per_pass} in Sect. \ref{subs_prec_flux}). In our model, we used $B=5$\,nT.

\subsubsection{The electron precipitation flux}
\label{subs_prec_flux}

In this section, we determine the electron precipitation flux, that is, the flux that reaches the atmosphere of Triton and therefore interacts with atmospheric species. This flux is different from the electron flux that was measured in the Neptune magnetosphere with the LECP instrument of Voyager 2 \citep{krimigis_hot_1989} (called magnetospheric flux in the following). The main problem here is that the transition from the magnetospheric flux to the precipitation flux is unknown. The total energy carried by the magnetospheric electrons with an energy higher than 20\,keV in the Neptune magnetosphere is 0.42\,erg.cm$^{-2}$.s$^{-1}$ \citep{krimigis_hot_1989,sittler_tritons_1996}. This flux is more than 250 times higher than the flux needed to explain the Triton thermospheric temperature, which is 1.6.10$^{-3}$\,erg.cm$^{-2}$.s$^{-1}$ \citep{broadfoot_ultraviolet_1989}. Moreover, the magnetospheric flux was measured at 12\,$R_{\text{N}}$ from Neptune \citep{krimigis_hot_1989}, which is not in the direct vicinity of Triton.

In our work, we determined the electron precipitation flux reaching the Triton atmosphere from the work of \citet{strobel_magnetospheric_1990} and \citet{sittler_tritons_1996}. We recall, however, that the steps we took to compute the precipitation flux from the magnetospheric flux may not result in a flux that is representative of the real precipitation flux at Triton. Only an orbiter measuring the magnetic environment of Triton over many orbits can reveal the precipitation flux at Triton. Here, we tried to determine the flux based on the few observations we have, physics-based considerations, and following what was assumed in previous studies.
\newline

The magnetospheric flux was taken from \citet{strobel_magnetospheric_1990}, based on LECP and PLS measurements from Voyager 2 near the Triton flyby. It corresponds to a flux $j(E)$ (in cm$^{-2}$.s$^{-1}$.sr$^{-1}$.keV$^{-1}$) of

\begin{align}
      j(E) &= 2.10^4 \times \left ( \frac{E}{28\,\text{keV}} \right )^{-2.7}
       \text{   for $E\geq$28\,keV} \label{eq_flux_S1990_1} \\
      j(E)& = 2.10^4 \times \left ( \frac{E}{28\,\text{keV}} \right )^{-0.7}
      \text{   for $E$ $<$ 28\, keV},
      \label{eq_flux_S1990_2}
\end{align}
where $E$ is the energy of the electrons in keV.

\begin{figure}[!h]
   \centering
   \includegraphics[width=0.65\textwidth]{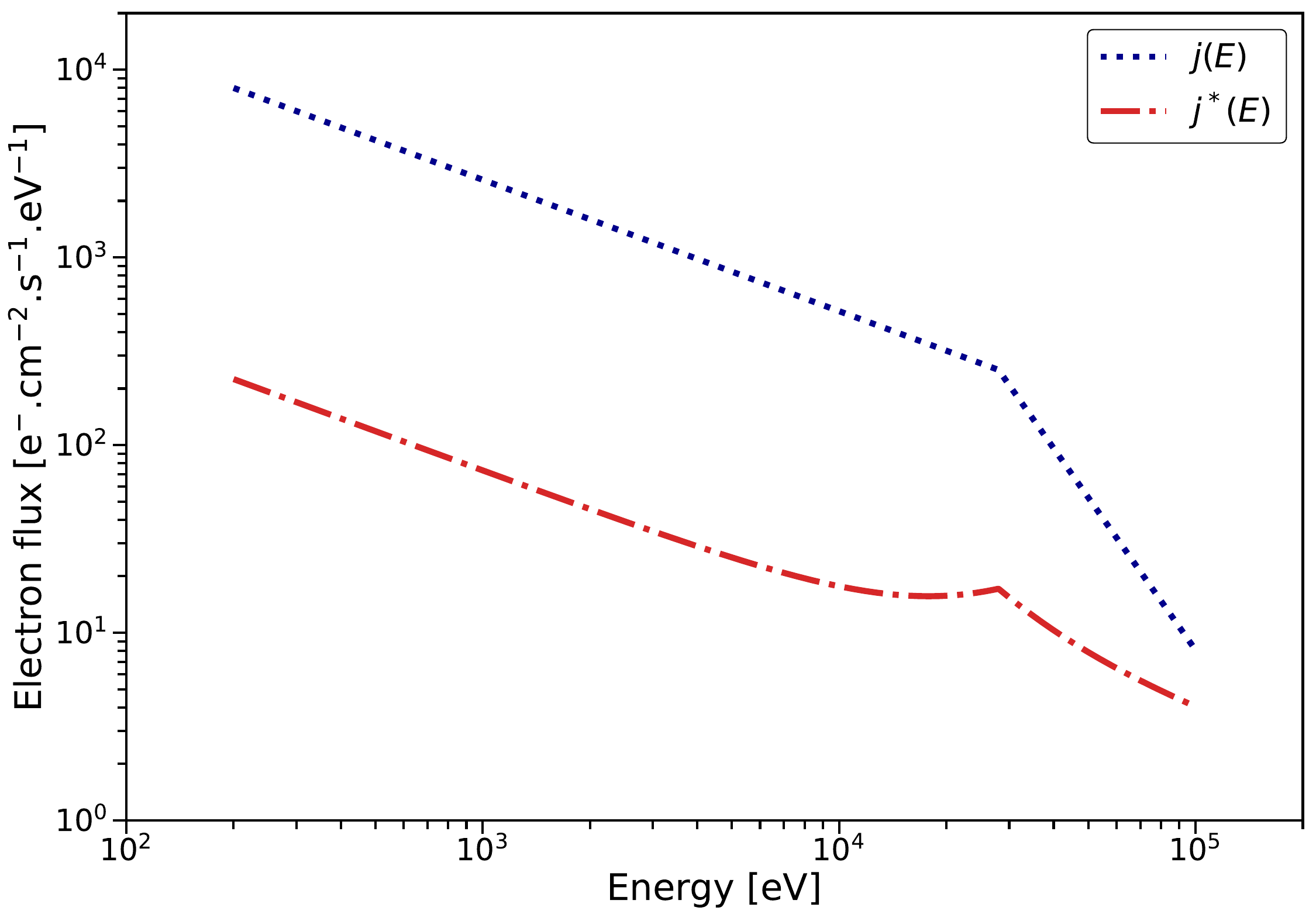}
   \caption{Comparison of the magnetospheric electron flux $j(E)$ from \citet{strobel_magnetospheric_1990} (dotted blue line), derived from Voyager 2 observations, and the modified flux $j^*(E)$ (dash-dotted red line), computed following the recommendations of \citet{sittler_tritons_1996}. The flux $j^*(E)$ is computed from the flux $j(E)$ multiplied by a precipitation probability $p(E)$.}
   \label{compar_fluxes_S1990_S&H96}
\end{figure}

\noindent According to \citet{sittler_tritons_1996}, not all the electrons contained in the magnetospheric flux reach the Triton atmosphere. Their results show that the higher the electron energy, the more likely the precipitation. As stated in their article, this contradicts \citet{summers_tritons_1991} (and therefore \citealt{krasnopolsky_temperature_1993,krasnopolsky_photochemistry_1995,strobel_tritons_1995,benne_photochemical_2022}), who inferred from the shift in the ionization profile of \citet{strobel_magnetospheric_1990} toward higher altitudes that the precipitation of high-energy electrons is inhibited.  More precisely, \citet{sittler_tritons_1996} computed that fewer than 3\% of the electrons with energies lower than 5 keV precipitate, fewer than 15\% of electrons with energies between 5 and 50 keV, and half of the higher-energy electrons.
Therefore, we modified the magnetospheric flux $j(E)$ from \citet{strobel_magnetospheric_1990} to account for these probabilities. We thus computed a flux $j^*(E)$, such that

\begin{equation}
   \label{eq_Jstar}
   j^*(E) = j(E)\times p(E),
\end{equation}
\noindent where $p(E)$ is the precipitation probability. We considered that 3 and 15\% of the electrons with energies of 5 and 50 keV precipitate, respectively, and 50\% of the electrons with the highest energy considered in TRANSPlanets, that is, 100 keV. To compute the precipitation probability $p(E)$ for the electrons with intermediate energies, we fit a second-order polynomial using these three values.
The $j(E)$ and $j^*(E)$ fluxes are plotted in Fig. \ref{compar_fluxes_S1990_S&H96}. 

Knowing $j^*$, we can estimate the power that is deposited by the electrons in the Triton atmosphere. \citet{strobel_magnetospheric_1990} claimed that the power is deposited by curvature drift. In this case, the deposited power $P_{\text{CD}}(E)$ is computed with formulas \eqref{power_deposited} and \eqref{drift_per_pass}, taken from \citet{strobel_magnetospheric_1990},

\begin{equation}
   P_{\text{CD}}(E) = j^*(E) \times \pi \times R_i \times \delta(E)
   \label{power_deposited},
\end{equation}
\noindent where $R_i$ is the radius of the ionopause, and $\delta$ is the inward drift per pass through the ionosphere, such as

\begin{equation}
   \delta = \frac{c m_e v_{//} \pi}{e B}  
   \label{drift_per_pass},
\end{equation}
\noindent with $c$ the speed of light, $m_e$ the electron mass, $v_{//}$ the electron speed parallel to the magnetic field, $B$ the strength of this field, and $e$ the elementary charge. 

 \citet{sittler_tritons_1996} defined the radius of the ionopause as $\frac{1}{\sigma n} \approx R_{i}$, with $\sigma$ the ion-neutral cross section, and $n$ the neutral number density. In our case, this gives $R_i \approx 2\,300$\, km, which nearly corresponds to their value of $\approx$ 2\,350\,km, while \citet{strobel_magnetospheric_1990} took $\approx$ 2\,000\,km. 

Using the flux $j$ in Eq. \eqref{power_deposited}, considering $B=8$\,nT in Eq. \eqref{drift_per_pass}, and using the flux between $E_{\text{min}}=200$\,eV and $E_{\text{max}}=10^5$\,eV, we find $P(E)=8.10^8$\,W.
This power can be compared to the total power carried by the electron flux, $E_{\text{flux}}$, which is computed as

\begin{equation}
   P_{\text{flux}} = 4\pi R_i^2\times 1,602176634.10^{-19}\times \int_{E_{\text{min}}}^{E_\text{max}} j(E)EdE ~~\text{[W]}.
\end{equation}
\noindent With $E_{\text{min}}=200$\,eV and $E_{\text{max}}=10^5$\,eV, we find $P_{\text{flux}}=3,4.10^{10}$\,W. 
Therefore, the power deposited by curvature drift is lower than the total power carried by the electron flux $j(E)$. 

The same can be found for $j^*$: The total power carried by the electrons is 3.8$\times$10$^{9}$\,W, whereas the power that should be deposited by curvature drift is only 1.8$\times$10$^8$\,W (using $B=5$\,nT, as computed in Sect. \ref{Trit_envrnmt}). Thus, to deposit the correct amount of power in the atmosphere, that is, to mimic a power deposition by curvature drift, we chose to renormalize the precipitation flux using Eq. \eqref{eq_renorm_flux}, 

\begin{equation}
   \label{eq_renorm_flux}
   j^*_{CD}(E) = j(E)\times p(E) \times \frac{P_{\text{CD}}(E)}{P_{\text{flux}}(E)}.
\end{equation}

\begin{figure}[!h]
   \centering
   \includegraphics[width=0.65\textwidth]{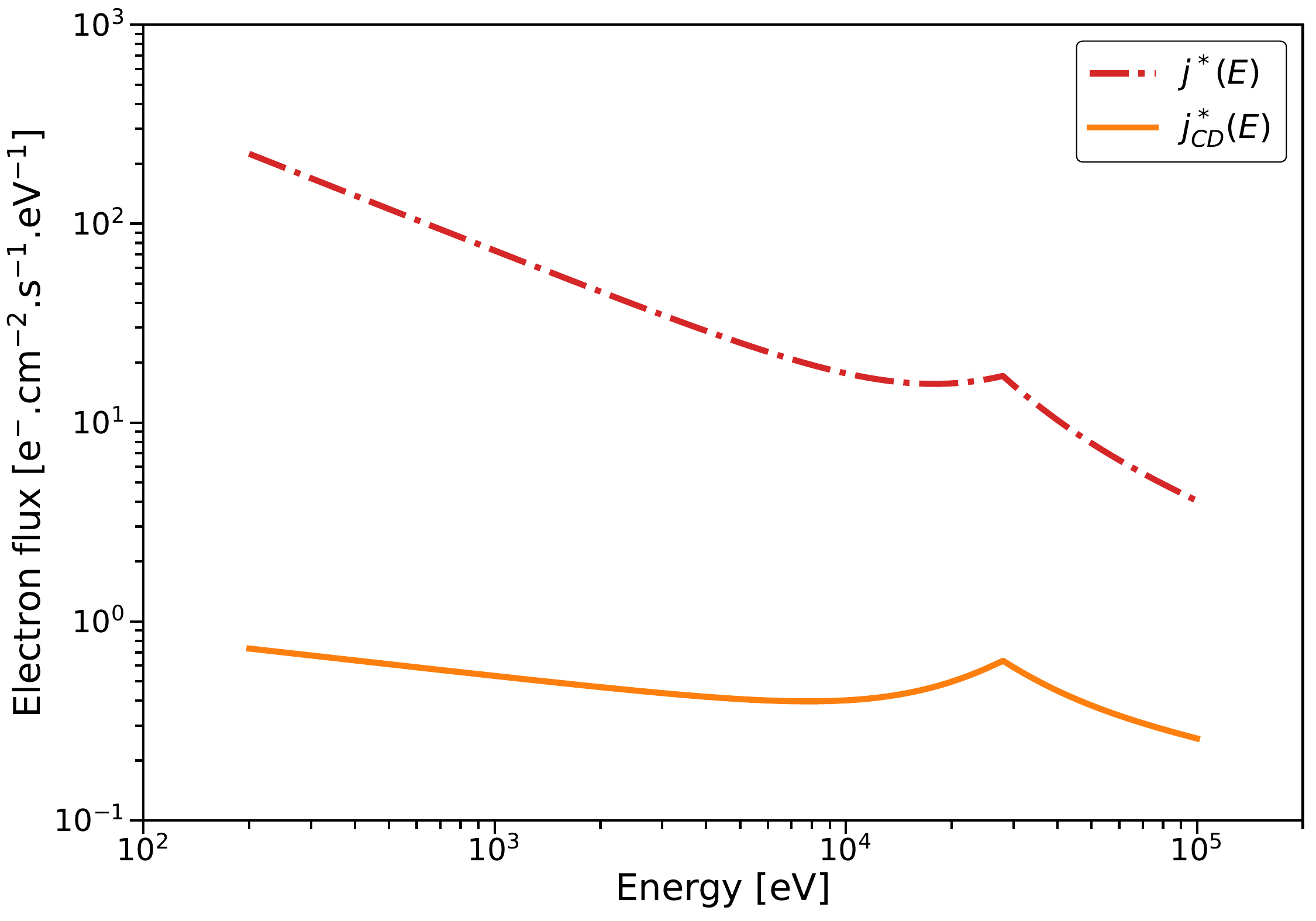}
   \caption{Comparison of the electron flux $j^*(E)$ (dash-dotted red line) derived from the flux $j(E)$ from \citet{strobel_magnetospheric_1990} and Eq. \eqref{eq_Jstar} following the results of \citet{sittler_tritons_1996}, and the renormalized flux $j^*_{CD}(E)$ (solid orange line). This flux is computed from $j^*(E)$ with Eq. \eqref{eq_renorm_flux}, so that the electrons deposit the correct amount of power in the Triton atmosphere, with the hypothesis that this power is deposited by curvature drift.}
   \label{compar_fluxes_S&H96_flux_renorm}
\end{figure}

\noindent The flux $j^*_{CD}(E)$ is compared to the flux $j^*(E)$ in Fig. \ref{compar_fluxes_S&H96_flux_renorm}.
Ultimately, the power carried by the flux $j^*_{CD}$ is 2.7$\times$10$^{-3}$\,erg.cm$^{-2}$.s$^{-1}$, which is more consistent with the value of 1.6$\times$10$^{-3}$\,erg.cm$^{-2}$.s$^{-1}$ computed by \citet{broadfoot_ultraviolet_1989} to explain the thermospheric temperature.

We clarify that we performed a test using the nonrenormalized flux $j^*$ in TRANSPlanets. In this case, all the power carried by the flux was classically deposited in the atmosphere, and therefore 3.8$\times$10$^9$\,W were deposited instead of 1.8$\times$10$^8$\,W. This resulted in high electron-impact ionization rates (see Fig. \ref{fig_prod_elec_fullbeans_annexe}), electron densities, and N number densities. These densities are inconsistent with Voyager 2 observations, as shown in Fig. \ref{fm_fullbeans_neutral+conc_ions_annexe} of Appendix \ref{Appendix_no_renorm}. This demonstrates the need to renormalize the precipitation flux so that the deposited power matches the deposit by curvature drift. 
\newline

Finally, we took the orbital scaling factor into account. \citet{strobel_magnetospheric_1990} multiplied the power deposited by curvature drift by $m_{orb}$ to allow for the intermittence of the precipitation. \citet{strobel_magnetospheric_1990} did not indicate whether the orbital scaling factor needed to be applied to the ionization profile, but this was done by \citet{krasnopolsky_photochemistry_1995}. In their model, they multiplied the ionization profile computed by \citet{strobel_magnetospheric_1990} by their adjusted orbital scaling factor of 0.162 to take into account the variability of the electron precipitation over one orbit.  In a similar way, we chose to multiply $j^*_{CD}$ by $m_{orb}$ in our model. We used $m_{orb} = 0.27$, as computed in Sect. \ref{subs_variable_params}, with our model of the Neptune-Triton system.

\subsection{Caveats}
\label{subs_Caveats}

Some assumptions and simplifications in our work have a significant impact on the results. They are listed below.
\begin{itemize}
   \item The magnetic field line is considered to be vertical. As mentioned in Sect. \ref{subs_fundamental_eqs}, this configuration is the most accurate when considering the projection of the axis parallel to the magnetic field line on the vertical axis. However, other configurations with inclined lines would be easily and accurately solved as well. Moreover, this allows modeling mean conditions, as was done for Titan in Dobrijevic et al. (2016). However, as considered in \citet{strobel_magnetospheric_1990} and \citet{sittler_tritons_1996}, the magnetic field lines are likely to drape themselves around the Triton atmosphere. Inclined lines change the thickness of the atmosphere that is crossed by the electrons, thus modifying the altitude at which they deposit their energy. Here, we lack knowledge of the magnetic environment of Triton. A systematic study of any field configuration is therefore meaningless, and we have to reduce our approach, keeping in mind that this is a severe limitation.
   \item As shown in the previous section, we used a magnetospheric electron flux that was measured at some distance from Triton to compute the precipitation flux used in TRANSPlanets. We also renormalized this flux to deposit the correct amount of power, considering that this power is deposited by curvature drift. Therefore, our input flux is dependent on the initial measurements, and we have no guarantees that this flux is representative of the real precipitation flux at the top of the Triton atmosphere. Until new in situ measurements of both the magnetospheric and ionospheric electron fluxes are made, however, we are limited by this assumption in our model. This can only be achieved with an orbiter performing several flybys of Triton for different magnetospheric configurations, and/or a probe/lander sent into the Triton atmosphere.
   \item We limited our calculations to 100\,keV, despite the measurements of higher-energy electrons in the Neptune magnetosphere and the results of \citet{sittler_tritons_1996} showing that higher-energy electrons are more likely to precipitate in the Triton atmosphere. We made this choice because TRANSPlanets solves the nonrelativistic Boltzmann equation and because our electron-impact cross sections for N$_2$ are only known until 100\,keV. There is currently no globally accepted way of taking into account relativistic electrons in electron transport models. Some solutions are to use the relativistic Boltzmann equation or relativistic cross sections. In this work, we chose to stay at relatively low energies to avoid modeling problems that are related to relativistic effects. 
   \item \citet{liuzzo_tritons_2021} modeled the interaction of Triton with the Neptune magnetosphere and predicted it to be complex. According to this model, the incident electron flux impacting the Triton ionosphere may strongly vary spatially and temporally. This variability is not taken into account in our model, as we chose to develop our approach based on Voyager 2 data, as was done in other photochemical models of the Triton atmosphere \citep{strobel_magnetospheric_1990,strobel_tritons_1995,krasnopolsky_photochemistry_1995,benne_photochemical_2022,krasnopolsky_tritons_2023}.
\end{itemize}

\subsection{Coupling TRANSPlanets with the photochemical model}
\label{couple_TRANS_PCM}

The remainder of the TRANSPlanets inputs were provided by our photochemical model, such as the atmospheric number density profiles of the neutral species and the thermal electrons. The altitude grid of the photochemical model was also used in TRANSPlanets. This grid was sampled with $H/5$ steps, where $H$ is the atmospheric scale height, giving 96 altitude levels between the surface and the top of the atmosphere (i.e., 1026\,km). We chose to consider five species in TRANSPlanets, namely N$_2$, N($^4$S), N($^2$D), C, and CO, as this allowed us to consider the three most abundant species in most of the atmosphere. The temperature profile for the neutral species was the same as in the photochemical model and was taken from the Triton-3 model of \citet{strobel_comparative_2017}. As in \citet{benne_photochemical_2022}, the electron temperature was taken to be equal to the neutral temperature. We adopted this hypothesis as no measurements of this physical parameter were made during the Voyager flyby. The electron temperature is likely to be higher than the neutral temperature, however, at least in the upper atmosphere. Using pressure balance arguments, \citet{sittler_tritons_1996} computed that the electron temperature at the ionopause should be between 308 and 1\,230\,K, depending on the hypotheses used. This is significantly higher than the neutral temperature in the thermosphere in our model, which is 92\,K. We ran some simulations with higher electron temperatures up to 1\,230\,K (see Sect. \ref{sect_sens_elec_temp}).
They produced a higher electron density because a higher electron temperature lowers the radiative and dissociative recombination rates. Therefore, to maintain a nominal electron profile consistent with Voyager 2 measurements, we would need to add another loss process for ions such as, for example, ion escape, as used in \citet{krasnopolsky_photochemistry_1995} and \citet{krasnopolsky_tritons_2023}.
However, we show in Sect. \ref{sect_sens_elec_temp} that the use of an electron temperature between 300 and 1\,230 K leads to electron density profiles that are not significantly different from those presented in \citet{tyler_voyager_1989} when we consider the chemical uncertainties and that the Tyler et al. profiles were given without any uncertainty. Thus, we chose to keep the electron temperature equal to the neutral temperature in our main model.
\newline

TRANSPlanets was used to compute the secondary ionization and dissociation of atmospheric species by a suprathermal electron flux. This flux consists of both precipitating electrons and photoelectrons. Therefore, the photoionization rates of the atmospheric species by EUV photons need to be computed first because they give the photoelectron flux as a function of energy. 
These photoionization rates were already computed by the photochemical model, with a high-resolution solar flux and high-resolution cross sections for N$_2$ and CO photolysis. However, these calculations only give the electron and ion production rates, thus losing all the information about the energy of the photoelectrons. In order to keep the energy dependence of the photoelectron flux, we calculated the photoionization rates within TRANSPlanets, adopting the solar flux and cross sections of the photochemical model.

We used the same solar flux in both codes, corresponding to maximum solar activity, as was the case at the time of the Voyager flyby. The flux had a resolution of 1 nm between 1 and 730 nm and was taken from \citet{thuillier_solar_2004}. 
Of the five species considered in TRANSPlanets, the photochemical model currently only considers the photoionization of N$_2$, N($^4$S), and CO. Therefore, TRANSPlanets only computes the primary ionization of these three species and the photodissociations of N$_2$ and CO. 
The maximum wavelength absorbed by at least one of these species being 163\,nm, we consequently restrained the calculations of TRANSPlanets to the 1-163\,nm interval. 
We also used the solar optical depths computed by the photochemical model in TRANSPlanets and a solar zenith angle of 50$^{\circ}$ in both codes, corresponding to mean global conditions at equinox\footnote{Using this solar zenith angle does not produce significant differences from the case using a classic solar zenith angle of 60$^{\circ}$, as the uncertainties on our results are large.}.
\newline

Finally, to be consistent between the photochemical model and TRANSPlanets, we considered the same electron-impact ionization and electron-impact dissociation reactions in both codes. These reactions are given in Table \Ref{electro_react_tab}. 
The chemical scheme we used is that of \citet{benne_photochemical_2022}, to which we added the last five reactions of Table \Ref{electro_react_tab}, as well as
the reactions (N($^4$S); N($^2$D)) + H$_2$CN $\longrightarrow$ (NH + HCN; $^3$CH$_2$ + N$_2$), whose rates were taken from \citet{hebrard_neutral_2012} with updated calculations.  

\begin{table}[!h]
     \begin{center}
        \begin{tabular}{l l l}
        \hline
        N$_2$ + e$^-$      & $\longrightarrow$ & N$_2^+$ + 2e$^-$           \\
        N$_2$ + e$^-$      & $\longrightarrow$ & N$^+$ + N($^2$D) + 2e$^-$  \\
        N$_2$ + e$^-$      & $\longrightarrow$ & N($^4$S) + N($^2$D) + e$^-$       \\
        N($^4$S) + e$^-$   & $\longrightarrow$ & N$^+$ + 2e$^-$             \\
        N($^2$D) + e$^-$   & $\longrightarrow$ & N$^+$ + 2e$^-$             \\
        C + e$^-$          & $\longrightarrow$ & C$^+$ + 2e$^-$             \\
        CO + e$^-$         & $\longrightarrow$ & CO$^+$ + 2e$^-$            \\
        CO + e$^-$         & $\longrightarrow$ & C + O($^3$P) + e$^-$             \\ 
        \hline      
        \end{tabular}   
        \caption[]{Electron-impact ionization and electron-impact dissociation reactions considered in the photochemical model and in TRANSPlanets.}
      \label{electro_react_tab} 
     \end{center}
\end{table}

\begin{figure}[!h]
   \centering
   \includegraphics[width=0.65\textwidth]{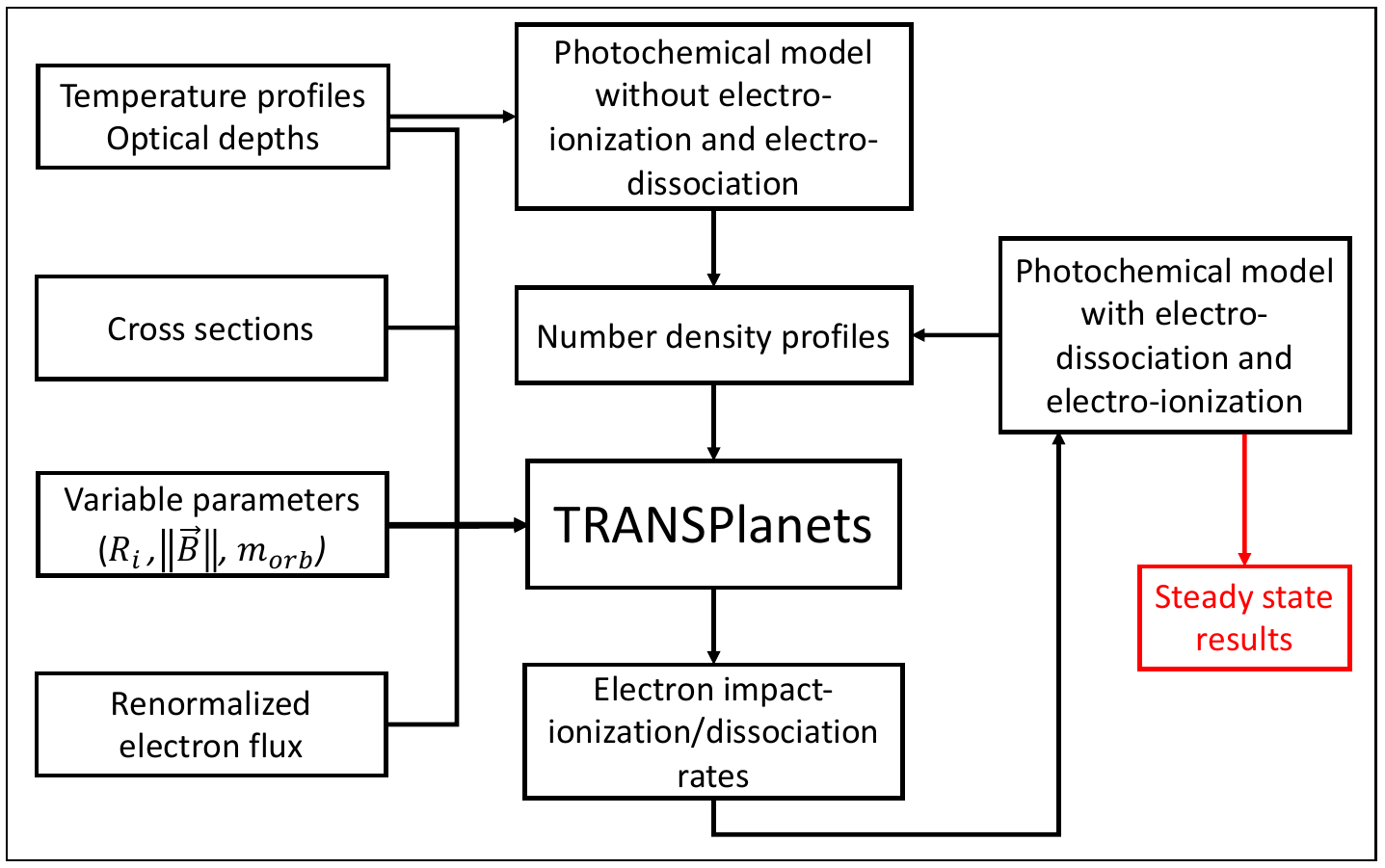}
   \caption{Method used to couple our photochemical model of the Triton atmosphere and the electron transport code TRANSPlanets. $R_i$ is the ionosphere radius, $\left \lVert \protect\overrightarrow{B} \right \rVert$ is the strength of the magnetic field at Triton, and $m_{orb}$ is the orbital scaling factor.
    }
   \label{scheme_TRANS}
\end{figure}

Our workflow was the following: We first ran the photochemical model without the electron-impact dissociation and electron-impact ionization reactions. Then, the N$_2$, N($^4$S), N($^2$D), C, CO and electron number density profiles, the neutral and electron temperature profiles, and the solar optical depths were used as inputs for TRANSPlanets. Then, TRANSPlanets computed the electron-impact ionization and electron-impact dissociation rates, which we subsequently used back in the photochemical model to calculate the atmospheric composition. We iterated this process until steady state was reached, that is, until the variations in the mole fractions between two successive iterations were small compared to the chemical uncertainties. This corresponded to three iterations. 
The method of the coupling of the two codes is presented in Fig. \ref{scheme_TRANS}.

\subsection{Global mean conditions}

\begin{figure}[!h]
   \centering
   \includegraphics[width=0.65\textwidth]{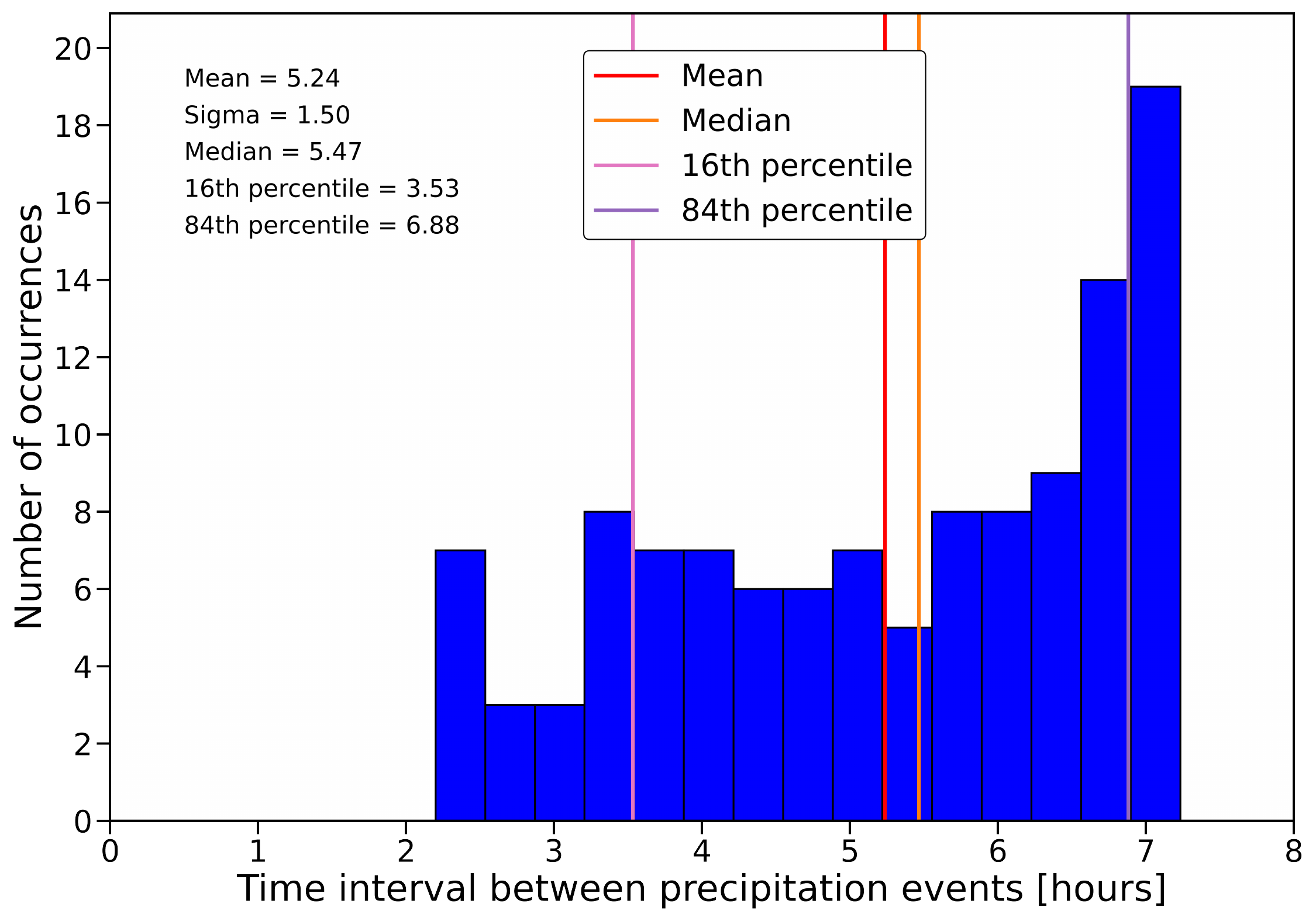}
   \caption{Histogram of the time interval between precipitation events computed with our model of the Neptune-Triton system. These values are computed over one period of variation in the magnetic field at Triton, i.e., 35 days, from August 21, 1989.
    }
   \label{Fig_Int_between_prec_events}
\end{figure}

We used our photochemical model to compute the steady-state composition of the Triton atmosphere. It thus ran over a long time interval, forcing us to consider global mean conditions. To verify whether global mean conditions were appropriate, we checked that the main ions of the Triton atmosphere had sufficiently long lifetimes to maintain an ionosphere between precipitation events. Based on previous modeling results \citep{majeed_ionosphere_1990,strobel_magnetospheric_1990,strobel_tritons_1995,krasnopolsky_photochemistry_1995,benne_photochemical_2022,krasnopolsky_tritons_2023}, the ionosphere may be mainly composed of atomic ions with slow recombination rates. 
We find that the lifetime of the main ion, C$^+$, varies between 4.8$\times$10$^6$\,s and 7.5$\times$10$^7$\,s in the ionosphere. The one of N$^+$ ranges between 3.9\,s at 204\,km and 5.6$\times$10$^5$\,s at 1026\,km, while the lifetime of H$^+$ lies between 3.1$\times$10$^6$\,s and 5.3$\times$10$^7$\,s. 
With our model of the Neptune-Triton system, we computed the time interval between precipitation events (see Fig. \ref{Fig_Int_between_prec_events}).
The mean value is 5.2 hours (1.9$\times$10$^4$\,s), and the time interval that occurs most frequently is 7.2 hours (2.6$\times$10$^4$\,s). This value is consistent with the time interval between each Neptune magnetic equator crossing by Triton  given in \citet{strobel_tritons_1995} and \citet{krasnopolsky_photochemistry_1995}, which is 7 hours. 

Therefore, the lifetimes of C$^+$ and H$^+$ are significantly longer that the time interval between precipitation events. The lifetime of N$^+$ is longer than 5.2 hours above 660\,km and longer than 7.2 hours above 676\,km. 
As a consequence, we find it unlikely that the abundances of these ions are significantly affected by the short-term variability in the magnetosphere. Global mean conditions for the photochemical model are thus valid.

After the steady-state composition of the atmosphere was determined, our model could compute the atmospheric composition under the specific conditions of the Voyager 2 flyby. However, we lack several critical inputs to do this, such as the electron precipitation flux and the magnetic field at Triton.

\section{Secondary ionization profiles}
\label{sec_e_prod}

In this section, we present the secondary electron production profile computed with TRANSPlanets. To do this, we used the flux $j^*_{CD}(E)$ presented in Fig. \ref{compar_fluxes_S&H96_flux_renorm}, an orbital scaling factor $m_{orb}=0.27$, and a magnetic field $B=5$\,nT. 
As done for Titan in \citet{dobrijevic_1d-coupled_2016}, we considered a vertical precipitation, even though magnetic field lines are expected to drape themselves around Triton \citep{strobel_magnetospheric_1990,sittler_tritons_1996} (see Sect. \ref{subs_Caveats}).
We emphasize that the following results were obtained considering secondary ionization and dissociation from both photoelectrons and magnetospheric electrons, which is not the case for the results presented in \citet{strobel_magnetospheric_1990}, where only the secondary ionization by magnetospheric electrons was considered. 

\begin{figure}[!h]
   \centering
   \includegraphics[width=0.65\textwidth]{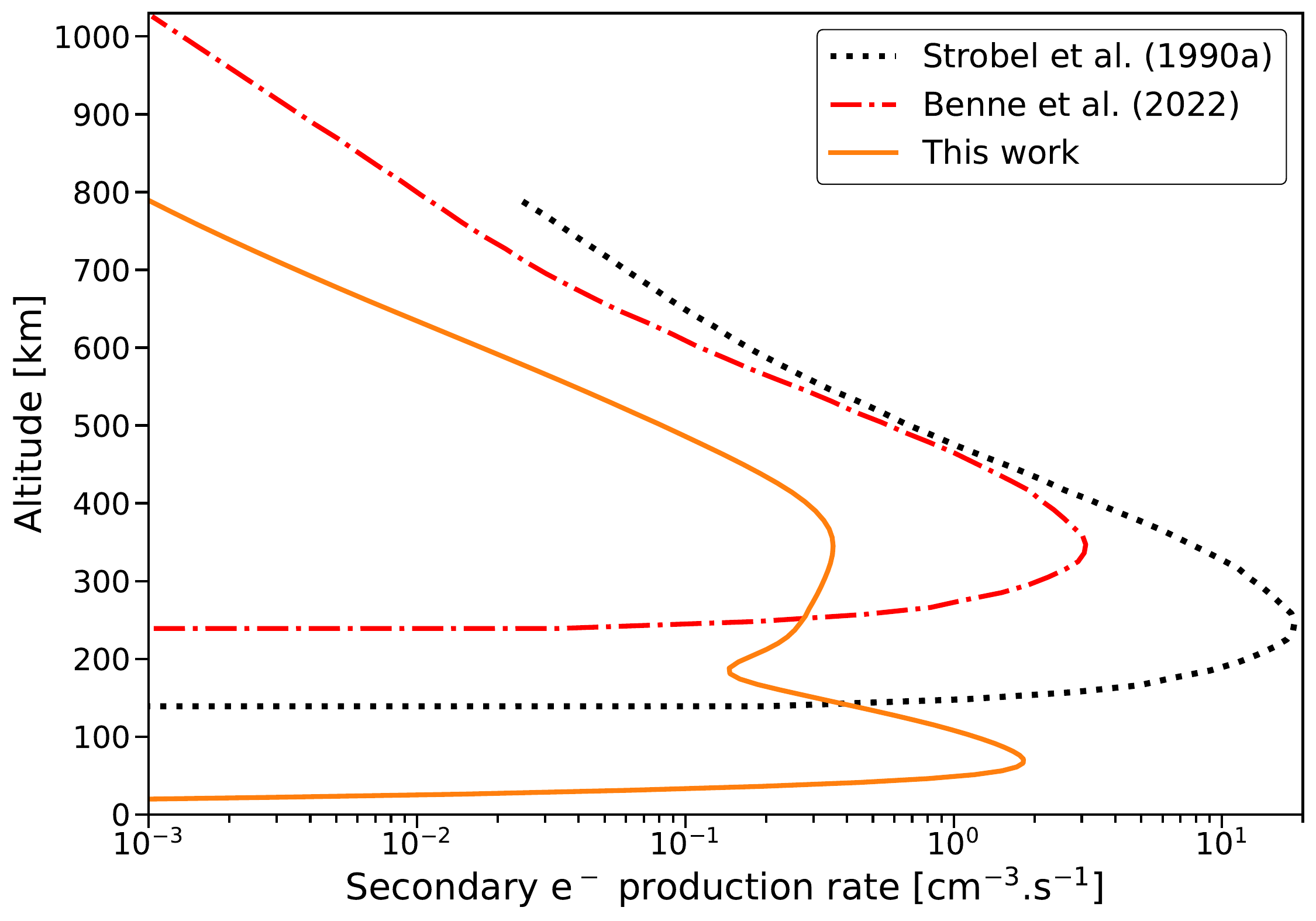}
   \caption{Secondary electron production profiles from \citet{strobel_magnetospheric_1990} (dotted black line), \citet{benne_photochemical_2022} (dash-dotted red line), and from this work (solid orange line).
    }
   \label{elec_prod}
\end{figure}

The secondary electron production profile obtained with our model is compared to those from \citet{benne_photochemical_2022} and \citet{strobel_magnetospheric_1990} in Fig. \ref{elec_prod}. 

Fig. \Ref{elec_prod} shows that we obtain a profile with two peaks. The first peak, around 400\,km, results from the ionization by photoelectrons. The second peak, at 71\,km, is due to electron-impact ionization by magnetospheric electrons. The maximum production rate of secondary electrons is 1.82\,cm$^{-3}$.s$^{-1}$. 
\newline

In addition to this profile, TRANSPlanets allowed us to obtain the reaction rates for every reaction of electron-impact ionization and electron-impact dissociation separately. This is the main difference between this work and the model of \citet{benne_photochemical_2022}. In the latter work, the electron-impact ionization and electron-impact dissociation rates were computed from the ionization profile shown in Fig. \ref{elec_prod} (dash-dotted red curve) and constant branching ratios from \citet{fox_electron_1988}:

\begin{table}[!h]
   \begin{tabular}{r l}
      0.8 & for the reaction N$_2$ + e$^-$ $\longrightarrow$ N$_2^+$ + 2e$^-$,\\
      0.2 & for N$_2$ + e$^-$ $\longrightarrow$ N$^+$ + N($^2$D) + 2e$^-$ \\
      0.6 & for N$_2$ + e$^-$ $\longrightarrow$ N($^4$S) + N($^2$D) + e$^-$.  
   \end{tabular}
\end{table}

\begin{figure*}
  \resizebox{\hsize}{!}
           {\includegraphics{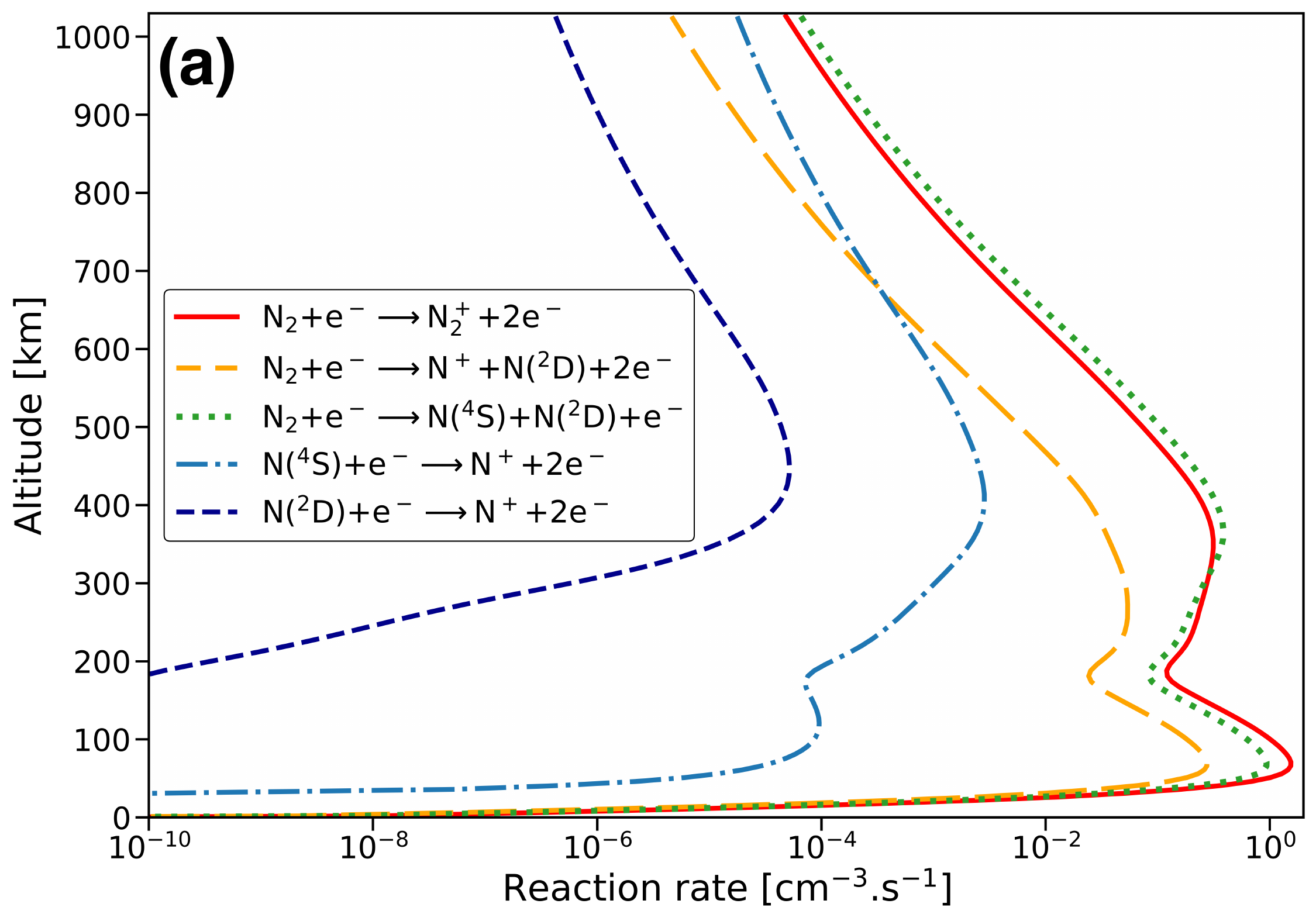}
           \includegraphics{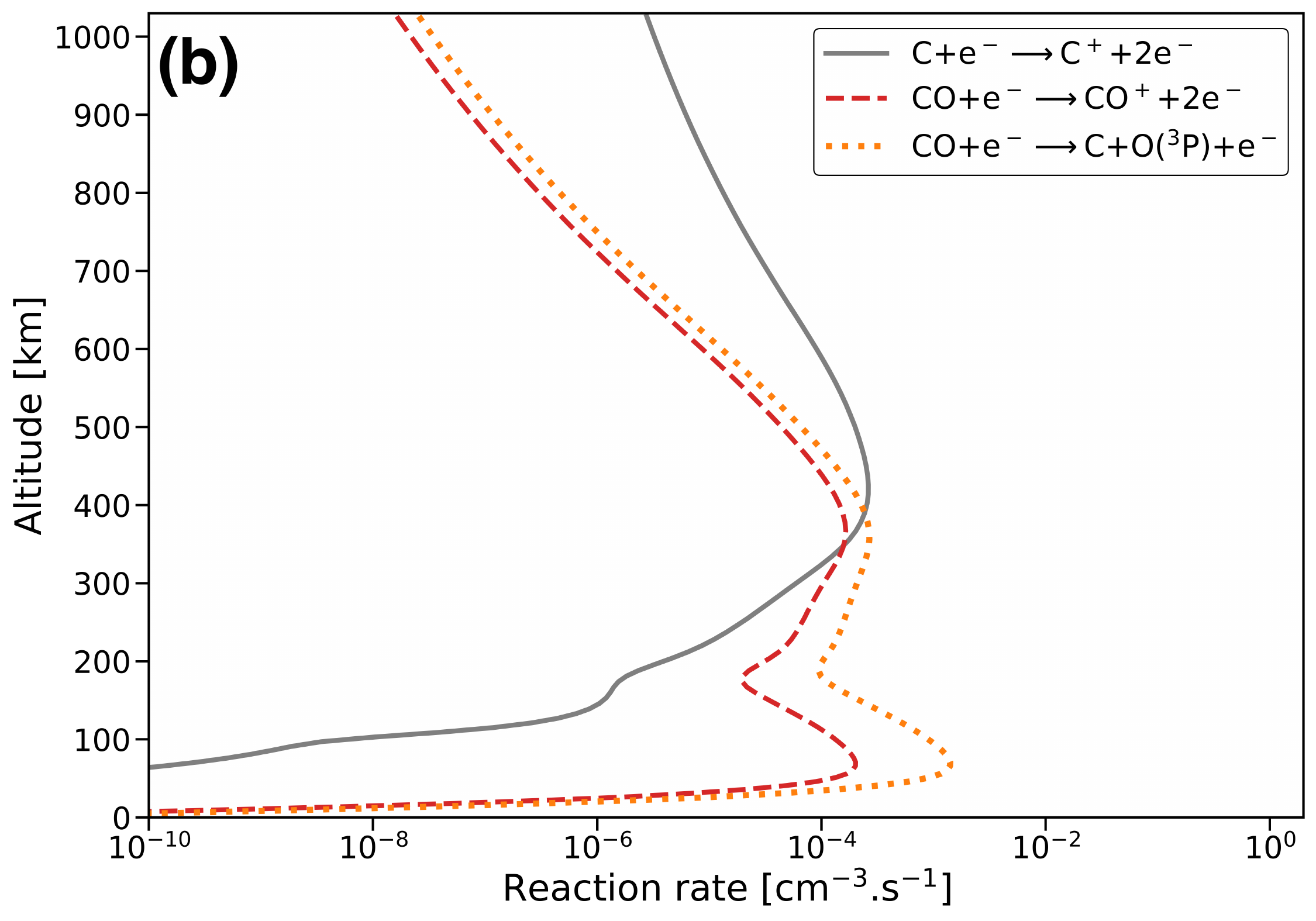}
           }
     \caption{Electron-impact ionisation and electron-impact dissociation rates computed with TRANSPlanets using the precipitation flux $j^*_{\text{CD}}$ multiplied by an orbital scaling factor of 0.27. \textbf{(a)} Reactions involving N$_2$, N($^4$S), and N($^2$D). \textbf{(b)} Reactions involving C and CO. 
             }
        \label{reac_rates_ME_TRANS}
\end{figure*}

\noindent Using TRANSPlanets, we did not need to use branching ratios because the reaction rates were computed independently for each reaction. Therefore, these profiles, one per reaction presented in Table \ref{electro_react_tab}, were used in the photochemical model. They are presented in Fig. \ref{reac_rates_ME_TRANS}. We describe the results from the photochemical model using these rates in the next section. 

\section{Results and discussion}
\label{sec_NomRes}

In the sections that follow, we compare some of the results from this work to results from a model that did not use TRANSPlanets, similar to \citet{benne_photochemical_2022}. This model is a slightly modified version of the model from \citet{benne_photochemical_2022}, in which we improved the way in which the interstellar medium flux is taken into account. This slightly modifies the nominal results presented in \citet{benne_photochemical_2022}, but the changes remain small compared to the chemical uncertainties. Moreover, the key chemical reactions and the main chemical pathways remain the same. Thus, we refer to this model as \citet{benne_photochemical_2022}. We recall that in this model, the electron-impact dissociation and electron-impact ionization rates were not computed with TRANSPlanets, but derived from the profile of \citet{strobel_magnetospheric_1990}, as described in Sect. \ref{sec_e_prod}.

As in \citet{benne_photochemical_2022}, we performed an evaluation of the uncertainties on the model results that are caused by the uncertainties in the chemical reaction rates. The method used here is the same as was presented in \citet{benne_photochemical_2022}. The Monte Carlo procedure was applied using 250 sets of the chemical reaction rates. 

\subsection{Ionization sources}

\begin{figure*}[!h]
   \resizebox{\hsize}{!}
      {
      \includegraphics[width=\hsize]{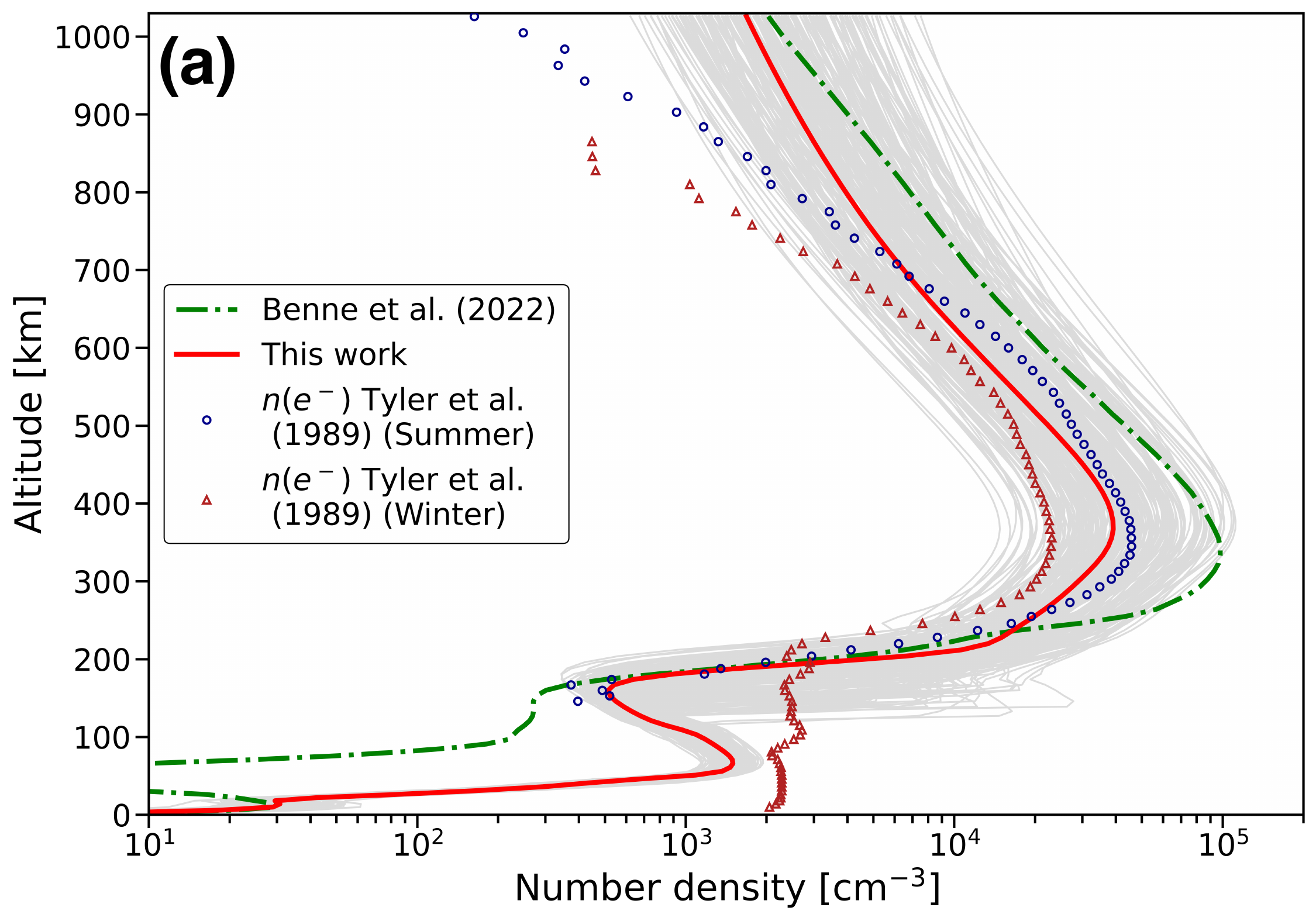}
      \includegraphics[width=\hsize]{Figures/histo_ratio_ioni250.pdf}
      }  
   \caption{Results from this work about the electron density and ionization processes. \textbf{(a)} Nominal electron density profiles from \cite{benne_photochemical_2022} (dash-dotted green line, using the electron production profile presented in Fig. \ref{elec_prod} and not TRANSPlanets) and from this work (red, using TRANSPlanets to compute the electron-impact ionization and electron-impact dissociation rates). The red triangles and blue circles give the electron density profiles measured by Voyager 2 at the winter and summer occultation points, respectively (data from \citealt{tyler_voyager_1989}). The gray profiles are the 250 electron profiles obtained from this work with the Monte Carlo procedure. \textbf{(b)} Histogram of the ratio of photoionization to electron-impact ionization obtained from the 250 runs of the Monte Carlo procedure.
    }
   \label{compar_elec_sol_flux}
\end{figure*}

The main result from this work is a significant decrease in the electron density in most of the atmosphere compared to \citet{benne_photochemical_2022}, as shown in panel (a) of Fig. \ref{compar_elec_sol_flux}, resulting from a decrease of 66\% in the electron-impact ionization rate, as listed in Table \ref{comp_int_col_rates_ME} of Appendix \ref{Appendix_comp_res}. For the nominal version of the model, ionization through photoionization is slightly more important than electron-impact ionization, with a ratio of 6/5. This is in contrast to the results of \citet{benne_photochemical_2022}, where this ratio was 3/8, similar to 1/2 found in \citet{krasnopolsky_photochemistry_1995}, even though we should underline that it is now close to unity. 
To confirm this result, we computed the photoionization to electron-impact ionization ratio for the 250 runs of the Monte Carlo procedure.
The histogram of the ratios is given in panel (b) of Fig. \ref{compar_elec_sol_flux}. We observe that electron-impact ionization is the main ionization source in the Triton atmosphere in only 16\% of the 250 runs of the Monte Carlo procedure, and the mean value of the photoionization to electron-impact ionization ratio is close to the nominal value of 6/5. This indicates that photoionization is slightly more efficient than electron-impact ionization in the Triton atmosphere. 
This result depends on the magnitude of the precipitation flux, however, and thus on the hypotheses we made to compute $j^*_{CD}(E)$. This point is discussed in Sect. \ref{subs_caveats_res}.

\subsection{Electron density}
\label{subs_Elec_dens}

With our model, the significant decrease in the electron-impact ionization rate results in a nominal electron peak density of 3.9$\times$10$^4$\,cm$^{-3}$ at 367\,km, which agrees well with the Voyager 2 observations of a peak of (3.5$\pm$1.0)$\times$10$^4$\,cm$^{-3}$ at (340-350)\,km \citep{tyler_voyager_1989,krasnopolsky_photochemistry_1995}. Therefore, the electron peak from this work is significantly smaller than that from \citet{benne_photochemical_2022}, where the peak electron density was 9.8$\times$10$^4$\,cm$^{-3}$ at 345\,km. Our results also agree better with the Voyager 2 observations than the results from \citet{strobel_tritons_1995} and \citet{krasnopolsky_photochemistry_1995}, who find their peaks at 275 and 320\,km, with peak number densities of 3.4$\times$10$^4$ and 3.5$\times$10$^4$\,cm$^{-3}$, respectively.

The electron density profile is strongly influenced by the reaction rate of the charge-exchange reaction between N$_2^+$ and C. This reaction is crucial for C$^+$ because it contributes 81\% of its integrated production. Since the main ion in our model is C$^+$, this reaction also impacts the electron density. The nominal rate constant of this reaction is 10$^{-10}$\,cm$^3$.s$^{-1}$, taken from the KIDA database \citep{wakelam_kinetic_2012}. The uncertainty factor for this reaction is $F(300 \text{ K})=3$. The rate of this reaction is thus significantly lower than the rate used in \citet{lyons_solar_1992}, who took 10$^{-9}$\,cm$^3$.s$^{-1}$, and explained the electron profiles with photoionization alone. This reaction is also a key uncertainty reaction, which means that it significantly affects the uncertainties on the model results. It is thus necessary to measure this reaction rate at temperatures representative of the Triton atmospheric conditions to improve Triton photochemical models. 

The results obtained with the Monte Carlo simulation are also consistent with the observations, as shown in panel (a) of Fig. \ref{compar_elec_sol_flux}. The electron density at 367\,km, the altitude of the nominal peak, is 4.2$\times$10$^4$\,cm$^{-3}$, with an uncertainty factor of 1.53. The altitude of the peak is (372$\pm$8)\,km at 1$\sigma$, which is slightly above the (340-350)\,km measured by Voyager 2 \citep{tyler_voyager_1989}. However, we note that the altitude of the electron peak depends on the inclination of the magnetic field line because it affects the atmospheric layers in which the electron-impact ionization and dissociation reactions take place. Thus, the altitude of the peak might change for a more inclined magnetic field line. For now, as shown in Fig. \ref{elec_prod}, the secondary electron production from electron-impact ionization by the precipitating electrons is maximum at 71\,km. A more inclined line should bring this peak to a higher altitude, which would shift the electron peak number density, possibly toward lower altitudes. 
We also note that in the end, the difference in the altitude of the electron peak may not be that significant because the data presented in \citet{tyler_voyager_1989} were given without any uncertainty. Voyager 2 RSS data was reanalyzed with modern tools by \citet{togni_voyager_2023}, who found that the electron peak may be located at higher altitudes than previously presented by \citet{tyler_voyager_1989}. 

\subsection{Results on the neutral atmosphere}

\begin{figure*}[!h]
           {\includegraphics[width=0.5\hsize]{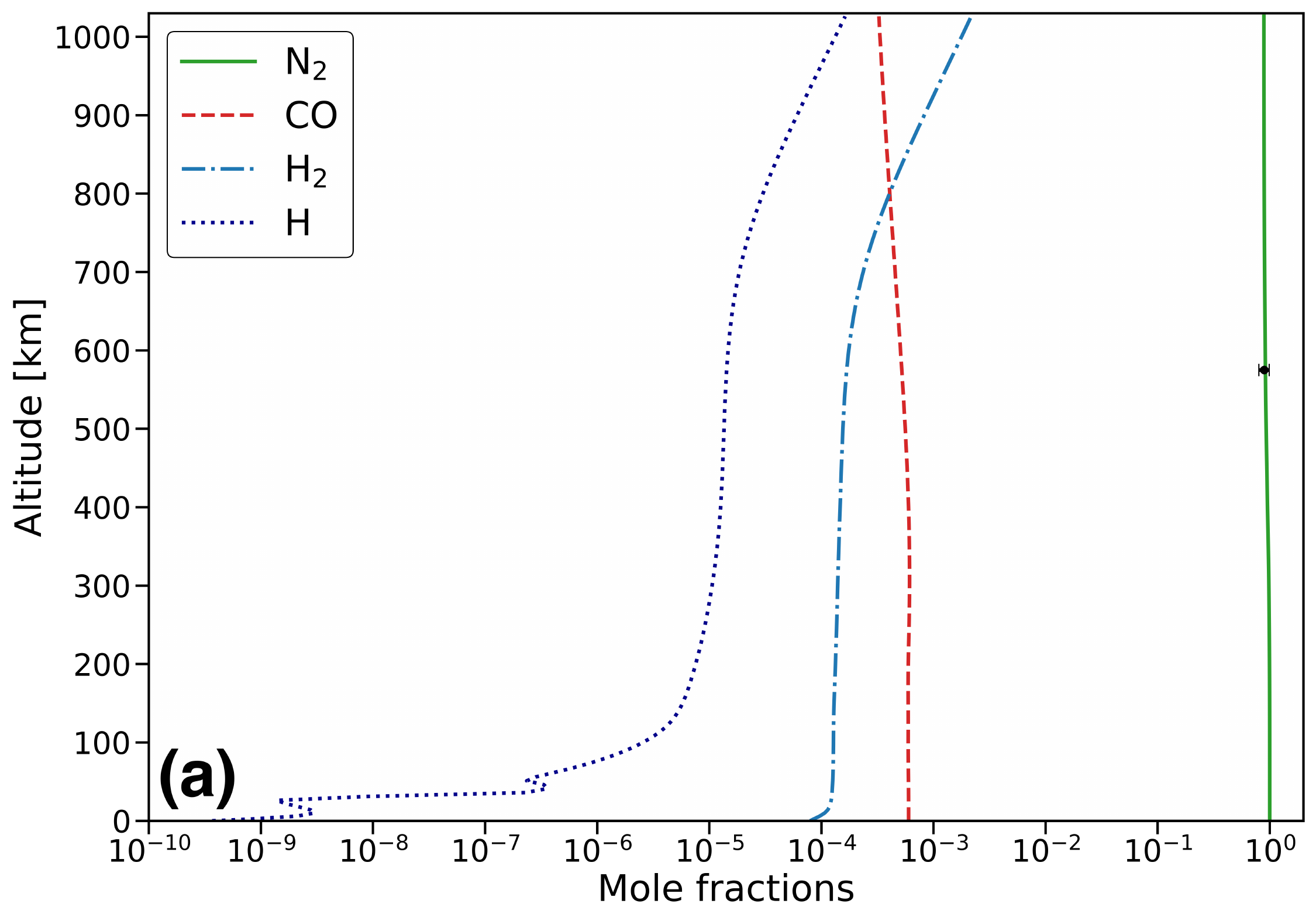}
           \includegraphics[width=0.5\hsize]{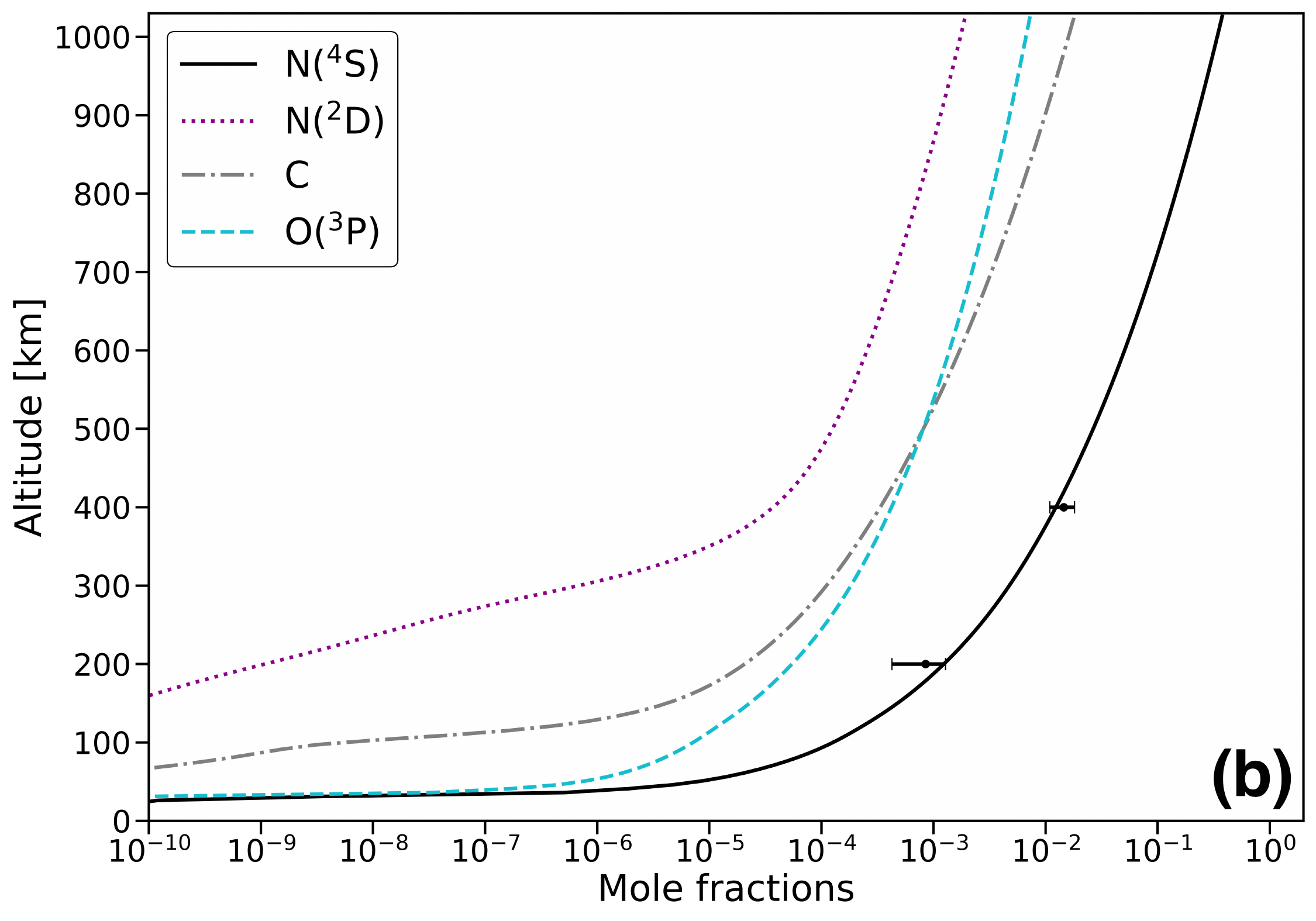}
           \includegraphics[width=0.5\hsize]{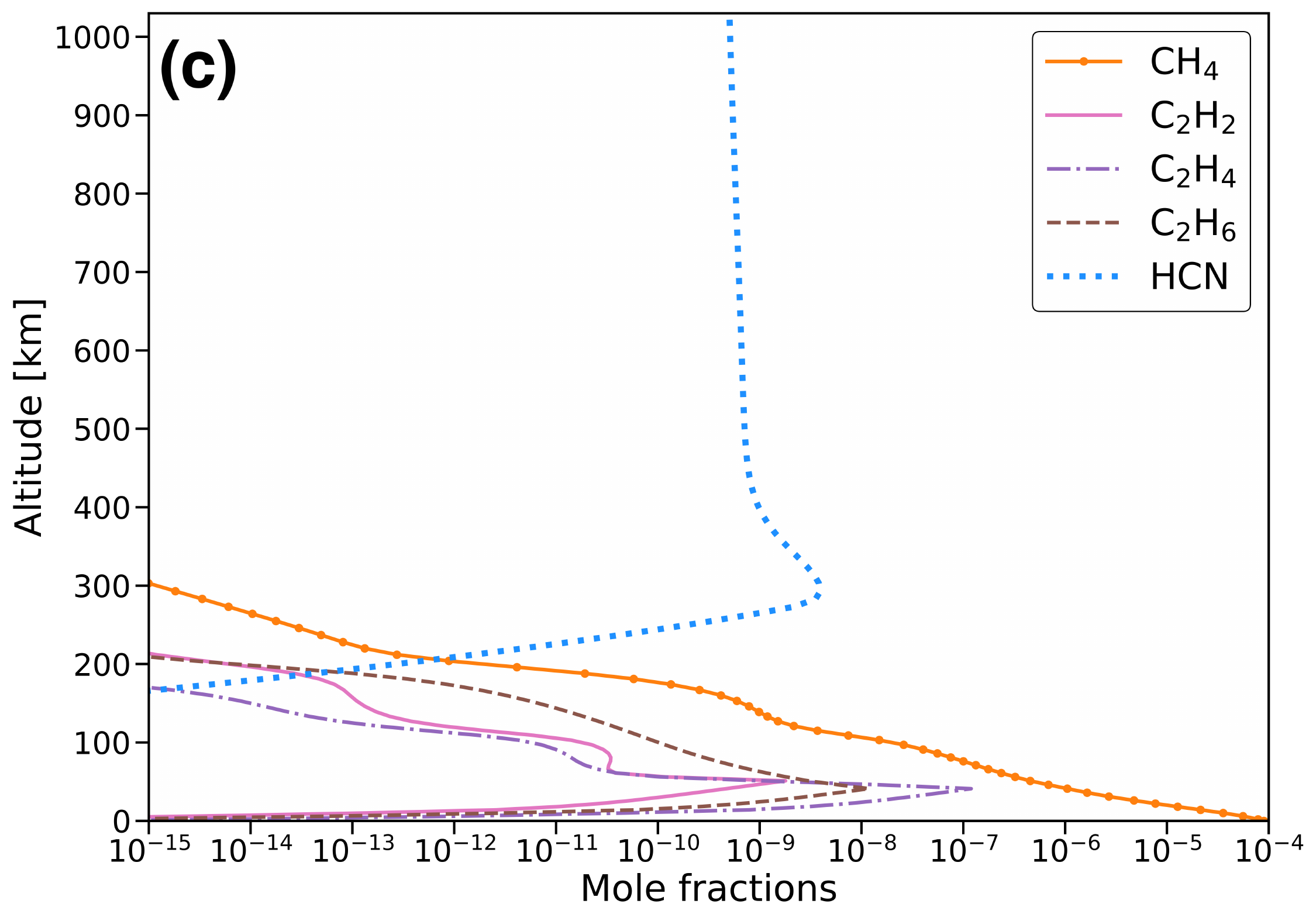}
           \includegraphics[width=0.5\hsize]{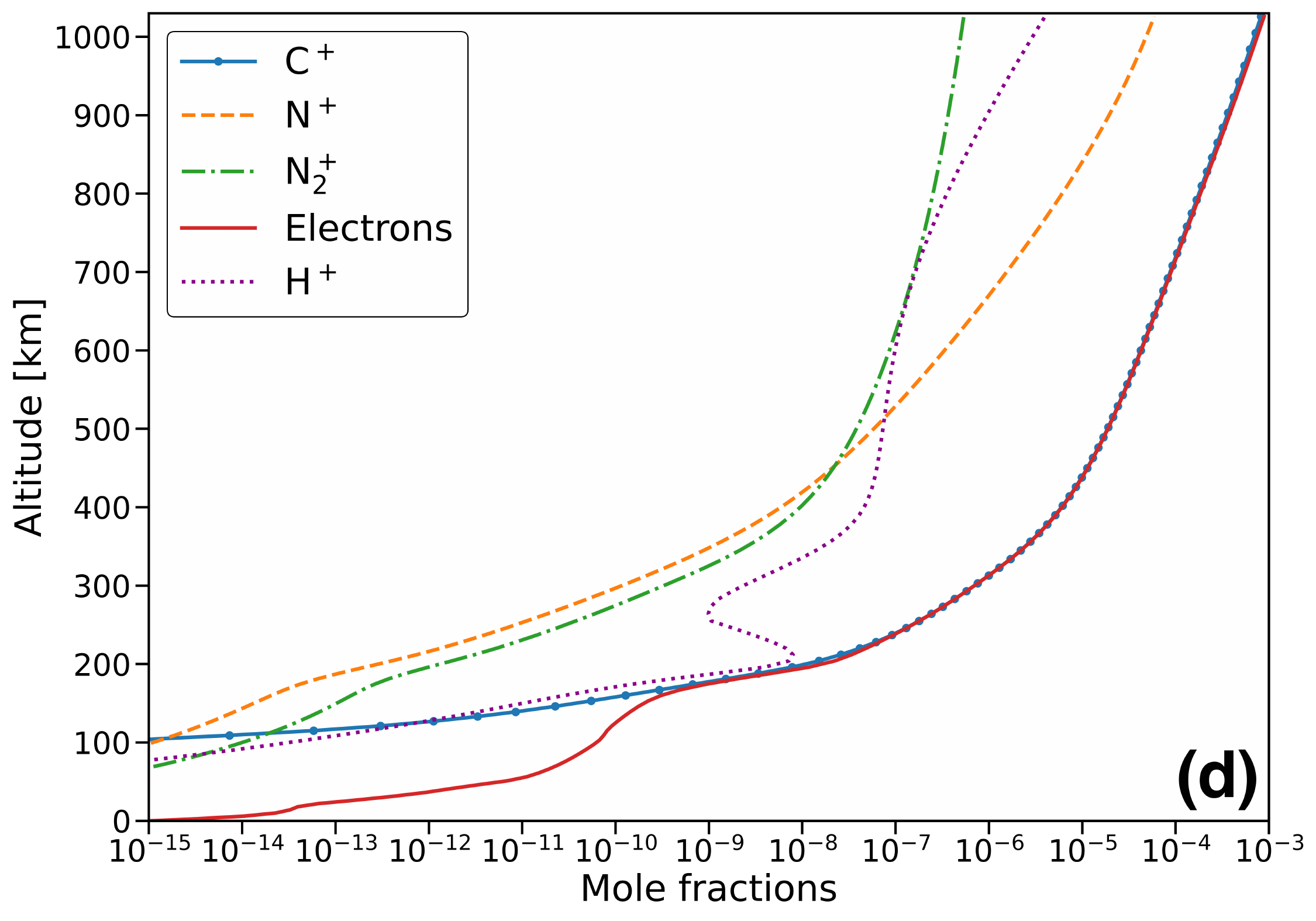}
           }
     \caption{Nominal simulated mole fraction profiles of the main species of the Triton atmosphere. (a) and (b): Main neutral species. (c) Main hydrocarbons and HCN. (d) Main ions. The bars on the N$_2$ and N($^4$S) profiles are the measurements from the UVS instrument of Voyager 2 \citep{broadfoot_ultraviolet_1989,krasnopolsky_temperature_1993}.
             }
        \label{compar_fm_main_species}
\end{figure*}

The nominal mole fraction profiles of the main species of the Triton atmosphere are given in Fig. \ref{compar_fm_main_species}. 

We obtain nominal N$_2$ and N number densities consistent with Voyager Ultraviolet Spectrometer (UVS) observations \citep{broadfoot_ultraviolet_1989,krasnopolsky_temperature_1993}, as shown in panels (a) and (b) of Fig. \ref{compar_fm_main_species}. We find a N$_2$ number density of 4.11$\times$10$^8$\,cm$^{-3}$ at 575\,km (with an uncertainty factor of 1.0), in agreement with (4$\pm$0.4)$\times$10$^8$\,cm$^{-3}$ from the observations.
For atomic nitrogen, we can compare our results to measurements at 400 and 200\,km, as given in \citet{krasnopolsky_temperature_1993,krasnopolsky_photochemistry_1995}. We note that for this species, we computed the uncertainty factor over the mole fraction of N($^4$S) at the corresponding levels, as its abundance is more than two orders of magnitude higher than that of N($^2$D) at 400\,km and six orders of magnitude higher at 200\,km. 
At 400\,km, we find a nominal number density of 8.38$\times$10$^7$\,cm$^{-3}$. Considering chemical uncertainties, we have a mean value of 6.42$\times$10$^7$\,cm$^{-3}$ with an uncertainty factor of 1.31. Thus, these values are consistent with the measurements of (1$\pm$0.25)$\times$10$^8$\,cm$^{-3}$ at this level.
In the 196-204\,km range, we have nominal values of 7.65-6.93$\times$10$^8$\,cm$^{-3}$ and mean values of 6.33-5.70$\times$10$^8$\,cm$^{-3}$ considering the chemical uncertainties, with uncertainty factors of 1.37 and 1.36, respectively, in agreement with (5$\pm$2.5)$\times$10$^8$\,cm$^{-3}$ at 200\,km measured from UVS.

An important outcome of this work is that we find nominal results consistent with the observations of Voyager 2 for the electron density profiles and the N$_2$ and N number densities, without having to apply ad hoc modifications to the electron-impact ionization profile (e.g., a shift toward higher altitudes), as was done in previous work \citep{summers_tritons_1991,krasnopolsky_photochemistry_1995,strobel_tritons_1995,benne_photochemical_2022}. In the latter article, the electron-impact ionization was computed using the ionization profile computed by \citet{strobel_magnetospheric_1990}. This profile was shifted by 100\,km to higher altitudes and multiplied by an orbital scaling factor, to take the intermittence of the precipitation into account. This orbital scaling factor was taken from \citet{krasnopolsky_temperature_1993}, who considered this factor as a free parameter of their model to be able to fit the observed electron density. In this work, we derived the orbital scaling factor from the modeling of the Neptune-Triton system by considering that this factor is the ratio of the time when Triton is near the magnetic equator (and therefore, where the precipitation occurs) over the total time considered. Therefore, this approach is only based on physical considerations.

\subsection{Model result uncertainties and main chemical pathways}

The study of the uncertainties on the model results shows us that despite the changes made in the modeling of the electron-impact ionization and dissociation reactions, the uncertainty factors remain large for the main species at the altitude where their uncertainty is maximum, even though we note some variations in comparison to \citet{benne_photochemical_2022}. This is shown in Table \ref{App_Table_F_uncert} of Appendix \ref{Appendix_comp_res}, which gives the uncertainty factors $F$ for all the main species of the Triton atmosphere at the level where the uncertainty on their mole fraction is maximum. We recall that the uncertainty factor $F$ was used to find the 1$\sigma$ interval around the mean mole fraction of a species, $\Bar{y_i}$, such that this interval is $\left [ \frac{\Bar{y_i}}{F},\Bar{y_i}\times F \right ]$ ($\Bar{y_i}$ is not necessarily equal to the nominal value of the mole fraction at the same altitude level). Finding large uncertainty factors is logical because the electron-impact ionization and dissociation of N$_2$ were only a small part of the key uncertainty reactions identified by \citet{benne_photochemical_2022}, and we used the same reaction rates for the other reactions. This confirms the need for new measurements at low temperatures for the key uncertainty reactions, as presented in \citet{benne_photochemical_2022}.

The main chemical pathways mostly remain the same between \citet{benne_photochemical_2022} and this work. We only observe some modifications of the contribution of the key chemical reactions to the chemical production and loss of the main atmospheric species (key chemical reactions are the crucial reactions producing or destroying a given species, as defined in \citealt{benne_photochemical_2022}). 
Because the main chemical pathways were already detailed in \citet{benne_photochemical_2022}, we only give the key chemical reactions from this work in Table \ref{comp_int_KCR} of Appendix \ref{Appendix_comp_res}.

\subsection{Caveats}
\label{subs_caveats_res}

The current model has some limitations, however. The first limitation is the geometry used for the electron precipitation: We considered it to be vertical, which led to a secondary electron production by these electrons that is maximum below 100\,km. The magnetic field lines are expected to drape themselves around Triton \citep{strobel_magnetospheric_1990,sittler_tritons_1996}, as is the case for Titan \citep{gronoff_ionization_2009-1,modolo_global_2008}. \citet{strobel_magnetospheric_1990} therefore considered a magnetic field line inclined by 9$^{\circ}$ and \citet{sittler_tritons_1996} worked with a horizontal line. The number of different configurations with a draping field is so large that a systematic study would be meaningless. We chose the configuration that was most accurate and prevented any additional uncertainty due to the projection toward the vertical. This configuration is found on Triton like any other and is similar to that used to study Titan in \citet{dobrijevic_1d-coupled_2016}. In the future, the study of other magnetic configurations as in \citet{gronoff_ionization_2009-1} will allow us to address how the energy is deposited in the atmosphere. For now, we did not compute the deposited energy flux, and we therefore cannot give any constraints on the heating of the atmosphere by magnetospheric electrons from our model. However, this was a critical argument supporting the electron precipitation hypothesis because the solar flux seemed too weak to explain the observed thermospheric temperature \citep{stevens_thermal_1992,krasnopolsky_temperature_1993,strobel_comparative_2017}. Again, as the electron-impact ionization and electron-impact dissociation reactions resulting from electron precipitation mostly take place below 100\,km, we expect the energy to be deposited in this area as well, which would certainly not allow us to explain the thermospheric temperature (if a secondary energy source is needed). 
However, considering a more inclined magnetic field line as described in Sect. \ref{subs_Elec_dens} should shift the region in which the electrons deposit their energy toward a higher altitude. Therefore, it should help to reproduce the thermospheric temperature.

Moreover, better constraints on the electron densities inside the Neptune magnetosphere and better constraints of the magnetic field strength at Triton depending on its position in the magnetosphere would allow us to improve our model significantly. This could be achieved with an orbiter equipped with a magnetometer and a plasma/energetic particles experiment that would make multiple flybys of Triton, as Cassini did for Titan. 

\subsection{Electron precipitation is still needed to explain the Voyager 2 observations}

We tested our photochemical model without considering electron precipitation and found an electron density profile consistent with the Voyager 2 observations presented in \citet{tyler_voyager_1989}, as shown in Fig. \ref{conc_ions_noME} of Appendix \ref{app_C}. 
However, even though a consistent electron density profile can be found without electron precipitation, it is more likely that this precipitation should be implemented in models of the Triton atmosphere \citep{lyons_solar_1992}. This additional source of energy is needed to explain the thermospheric temperature according to \citet{stevens_thermal_1992} and \citet{krasnopolsky_temperature_1993}. In addition, it is needed to obtain a number density of atomic nitrogen at 400\,km that matches the data presented in \citet{krasnopolsky_temperature_1993}.


\section{Sensitivity study of the electron temperature}
\label{sect_sens_elec_temp}

As explained in previous sections, the electron temperature $T_e$ was not explicitly provided in the majority of the articles about the photochemistry of the Triton atmosphere because no measurements of this physical quantity were performed during the Voyager 2 flyby. We therefore used an electron temperature equal to the neutral temperature in our work. Because the atmosphere of Triton is tenuous, however, we can expect $T_e$ to be higher than the neutral temperature $T_n$. Using various hypotheses to compute the pressure balance, \citet{sittler_tritons_1996} found that $T_e$ at the Triton ionopause should be between 308 and 1\,230\,K. These values are higher than the maximum neutral temperature reached in the thermosphere, which is 102$\pm$3 K according to \citealt{krasnopolsky_temperature_1993}. In their photochemical model, \citet{krasnopolsky_tritons_2023} assumed $T_e=300$\,K, but did not justify this choice.

\begin{figure}[!h]
   \centering
   \includegraphics[width=0.65\textwidth]{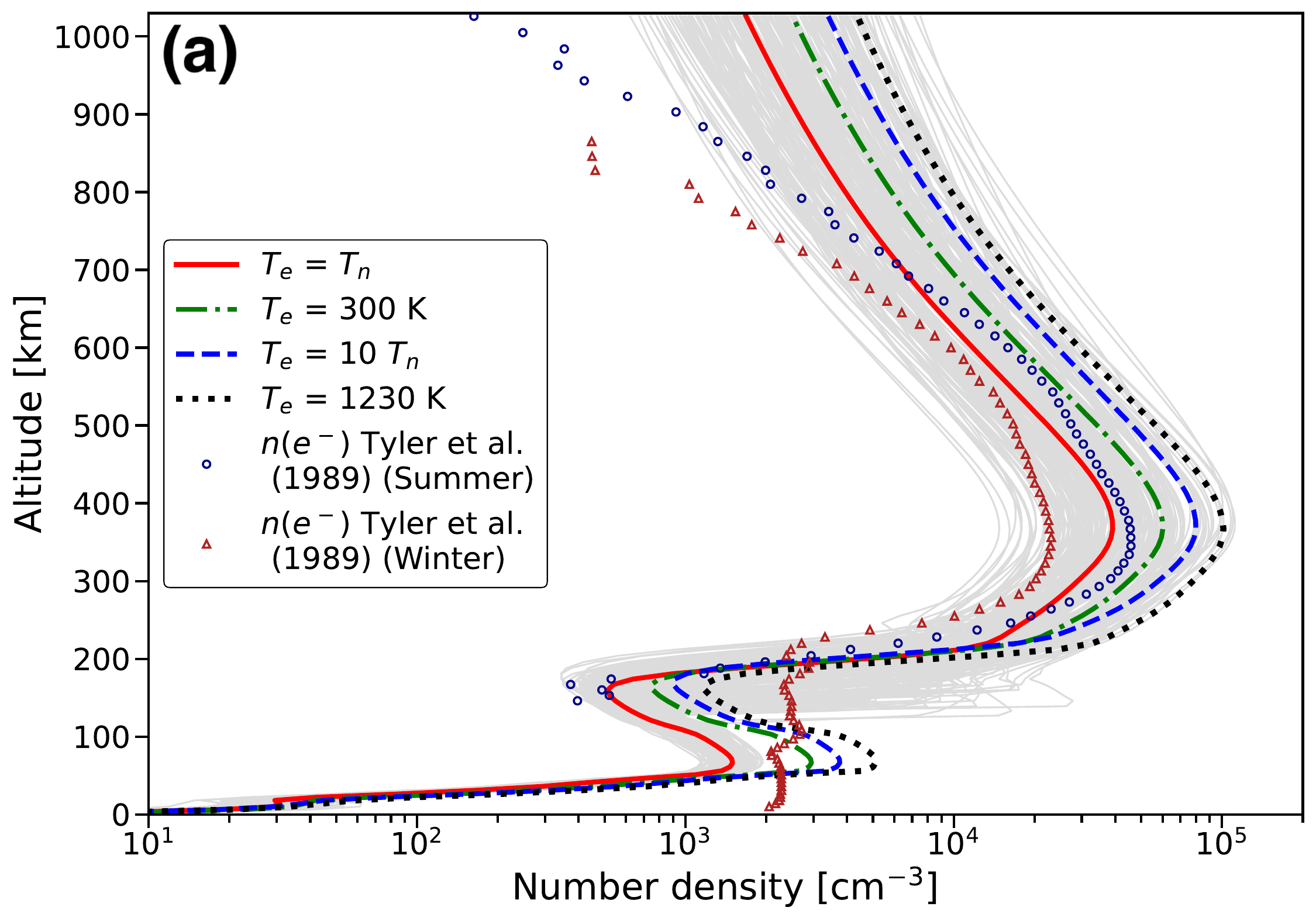}
   \caption{Nominal electron density profiles for $T_e=T_n$ (solid red line), $T_e=300$\,K (dash-dotted green line), $T_e=10\times T_n$ (dashed blue line) and $T_e=$1\,230\,K (dotted black line). The gray profiles are the electron density profiles obtained considering chemical uncertainties for the $T_e=T_n$ case. The red triangles and blue circles give the electron density profiles measured during the Voyager 2 flyby from \citet{tyler_voyager_1989}.
    }
   \label{compar_elec_dens_Te}
\end{figure}

As the electron temperature remains unknown, we performed multiple simulations with electron temperatures higher than the neutral temperature. Based on the results of \citet{sittler_tritons_1996}, our three main simulations considered $T_e=$\,300, 10$\times T_n$ (resulting in a maximum electron temperature of 920\,K) and 1\,230\,K. The results from these simulations are shown in Fig. \ref{compar_elec_dens_Te} and Table \ref{Tab_compar_elec_peak_Te}. 

\begin{table}[!h]
   \centering
   \begin{tabular}{@{}cccc@{}}
   \hline
   $T_e$    & \begin{tabular}[c]{@{}c@{}}Max $e^-$ density\\ {(}cm$^{-3}${)}\end{tabular} & \begin{tabular}[c]{@{}c@{}}Max $e^-$ \\ density\\ at 1-$\sigma$ {(}cm$^{-3}${)}\end{tabular} & \begin{tabular}[c]{@{}c@{}}$e^-$ density\\ peak {(}km{)}\end{tabular} \\ \hline
   $T_n$    & 4.2$\times$10$^4$ & [2.7-6.4].10$^4$  & 367    \\
   300 K    & 6.0$\times$10$^4$ & [3.9-9.2].10$^4$  & 367    \\
   10 $T_n$ & 8.0$\times$10$^4$ & [5.2-12.2].10$^4$  & 367   \\
   1\,230 K & 1.0$\times$10$^5$ & [6.5-15.3].10$^4$  & 367   \\ \hline
   \end{tabular}
   \caption[]{Comparison of the maximum electron density and its altitude for various cases using a different electron temperature profile. \\ The 1$\sigma$ interval is computed with the uncertainty factor of 1.53 that was calculated at the electron peak in Sect. \ref{subs_Elec_dens} for the $T_e=T_n$ case.}
   \label{Tab_compar_elec_peak_Te}
\end{table}

A higher electron temperature logically leads to a higher electron peak density because a higher electron temperature lowers the rates of dissociative and radiative recombination reactions. We observe in Table \ref{Tab_compar_elec_peak_Te} that the simulation with $T_e=300$\,K is consistent with Voyager 2 data from \citet{tyler_voyager_1989} at 1$\sigma$, which gives a maximum electron density of (3.5$\pm$1).10$^4$\,cm$^{-3}$. The simulations with $T_e=10\,T_n$ and 1\,230\,K give marginally higher electron peak densities at 1$\sigma$ because the values from \citet{tyler_voyager_1989} are given without any uncertainties. Moreover, the nominal electron density profiles remain in the uncertainty range of our reference case where $T_e=T_n$.

Therefore, increasing the electron temperature in our model leads to a higher electron density, but this increase is not significant considering the chemical uncertainties and that the profiles from \citet{tyler_voyager_1989} are given without uncertainties. In addition, recent reprocessing of Voyager 2 data by \citet{togni_voyager_2023} indicates that the electron density measured during the flyby is likely to be higher than what is presented in \citet{tyler_voyager_1989}. To make our nominal electron density profiles consistent with the profiles from \citet{tyler_voyager_1989}, we could also have added another unconstrained loss process in our model such as, for example, ion escape at the top of the atmosphere, as was done in \citet{krasnopolsky_photochemistry_1995,krasnopolsky_tritons_2023}. 

\section{Conclusion}
\label{conclu}

We presented a model that was improved over the model of \citet{benne_photochemical_2022} to better account for the interaction between the Triton atmosphere and the Neptune magnetosphere. To do this, we first modeled the Neptune-Triton system to study the variation in the magnetic environment of Triton depending on its position in the Neptune magnetosphere. The $L$-shell parameter of Triton varies strongly with time from 14 to more than 100. Thus, Triton is not always near the magnetic equator of Neptune, where the magnetospheric electrons are located. Therefore, we computed the time ratio when Triton is in this area, called the orbital scaling factor, and applied it to compute the mean precipitation, as done in \citet{strobel_magnetospheric_1990}. The initial input precipitation flux was taken from the latter article, but was then modified following the work of \citet{sittler_tritons_1996}, in which the electron precipitation rate depends on their energy.
We also computed the mean magnetic field using OTD models of the Neptune magnetic field that were developed after the flyby of Voyager 2. This gave a mean field of 5\,nT, slightly lower than the value of 8\,nT used by \citet{strobel_magnetospheric_1990} and \citet{sittler_tritons_1996}. This difference impacts the energy deposited by the electrons. We considered that the electron flux propagated vertically, that is, perpendicular to the atmospheric layers. 
\newline

With these inputs, we used the electron transport model TRANSPlanets to compute the electron-impact ionization and electron-impact dissociation rates. TRANSPlanets is a modified version of the TRANS model that has been used to study electron transport in various planetary atmospheres \citet{gronoff_ionization_2009}.
These rates were then used in our photochemical model after coupling it with TRANSPlanets. 
This significantly changed the results of the photochemical model compared to the case where the electron-impact ionization and electron-impact dissociation rates were computed as in \citet{benne_photochemical_2022}. The electron impact-ionization and dissociation integrated column rates decreased by 66\% between the two models, thus affecting the atmospheric composition. Consequently, electron-impact ionization is no longer the dominant ionization source in the Triton atmosphere because its rate is slightly lower than the rate of photoionization. This is opposite to what was found in the previous photochemical models of the Triton atmosphere \citep{strobel_tritons_1995,krasnopolsky_photochemistry_1995,benne_photochemical_2022}. However, we note that this result depends on the input flux used in TRANSPlanets, which depends on some of our assumptions because we lack knowledge of the electron fluxes at Triton, the time variation of its magnetic environment, and of the magnetosphere-atmosphere interaction. Maybe more importantly, these results also depend on the shape of the magnetic field around Triton. We made the strong assumption of a vertical magnetic field line. Considering the multiplicity of the different configurations of a satellite orbiting a complex magnetized planet, there is no doubt that this configuration may be found, and no doubt either that it is not a general rule. Therefore, this work must be considered as a first approach, showing that we are now able to address a detailed and deep study of the upper atmosphere of Triton. Observations are still missing, however. We strongly need a dedicated mission to study the Neptune magnetosphere in depth, with several flybys of Triton when it is at different places in the magnetosphere. 

Even though the atmospheric composition was altered, the main chemical pathways remained the same as in \citet{benne_photochemical_2022}. The changes come from the varying contribution of the different key chemical reactions. 
With this model, we find a nominal electron peak number density consistent with the Voyager observations presented in \citet{tyler_voyager_1989}, at an altitude that is roughly consistent with observations.
The nominal number densities of N$_2$ and N are also consistent with the observations reported in \citet{broadfoot_ultraviolet_1989,krasnopolsky_temperature_1993,krasnopolsky_photochemistry_1995}. We performed a Monte Carlo simulation with 250 runs of the model by varying the chemical reaction rates. We found that some uncertainty factors were lower in this work compared to \citet{benne_photochemical_2022}, but the overall uncertainties are still large. 
We also showed that using increased electron temperature leads to higher electron densities, but these densities remain consistent with the Voyager 2 observations presented in \citet{tyler_voyager_1989} when we consider the chemical uncertainties and that the profiles from \citet{tyler_voyager_1989} did not display any uncertainty.
\newline

The current model can be improved in the following ways. The possibility of using a horizontal magnetic field line could have a significant impact on the results by shifting the electron-impact ionization profile resulting from the electron precipitation upward. This case would also model the Triton-magnetosphere interaction better because it is thought that the magnetic field lines drape themselves around Triton \citep{strobel_magnetospheric_1990,sittler_tritons_1996}. The computation of the heat deposited by the electron precipitation would also help to constrain the importance of the geometry of the magnetic field lines in the upper atmosphere of Triton because magnetospheric electrons may explain the thermospheric temperature \citep{stevens_thermal_1992,krasnopolsky_temperature_1993,strobel_comparative_2017}. 
This model would also benefit from a better knowledge of the Neptune magnetosphere, such as the electron number density depending on the $L$-shell to better model the input precipitation flux. The acquisition of this type of data requires an orbiter in the Neptunian system. 
\section*{Acknowledgements}
B. Benne, M. Dobrijevic and T. Cavalié. acknowledge funding from CNES.
B. Benne, M. Dobrijevic, T. Cavalié and K.M. Hickson acknowledge funding from the ‘‘Programme National Planétologie’’ (PNP) of CNRS/INSU.
K.M. Hickson. acknowledges support from the French program ‘‘Physique et Chimie du Milieu Interstellaire’’ (PCMI) of the CNRS/INSU with the INC/INP co-funded by the CEA and CNES.  
\newpage 

\begin{appendix} 
  
  \section{Results obtained with a not renormalized precipitation flux}
  \label{Appendix_no_renorm}
  
\begin{figure}[!h]
     \centering
              {\includegraphics[width=0.65\textwidth]{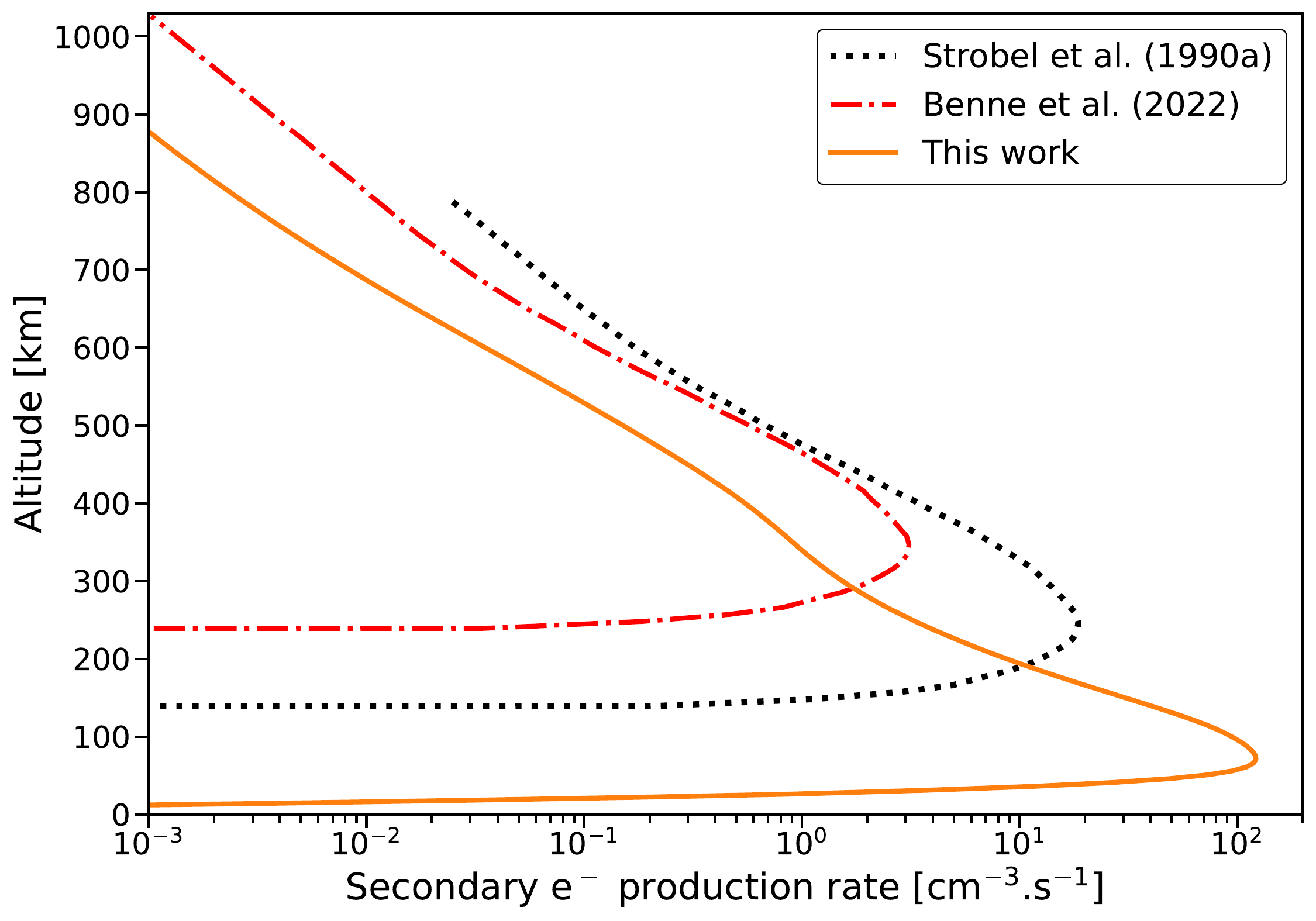}
              }
        \caption{Secondary electron production profile computed with TRANSPlanets using a not renormalized precipitation flux $j^*(E)$.
                }
           \label{fig_prod_elec_fullbeans_annexe}
  \end{figure}

\begin{figure*}[!h]
              {\includegraphics[width=0.5\hsize]{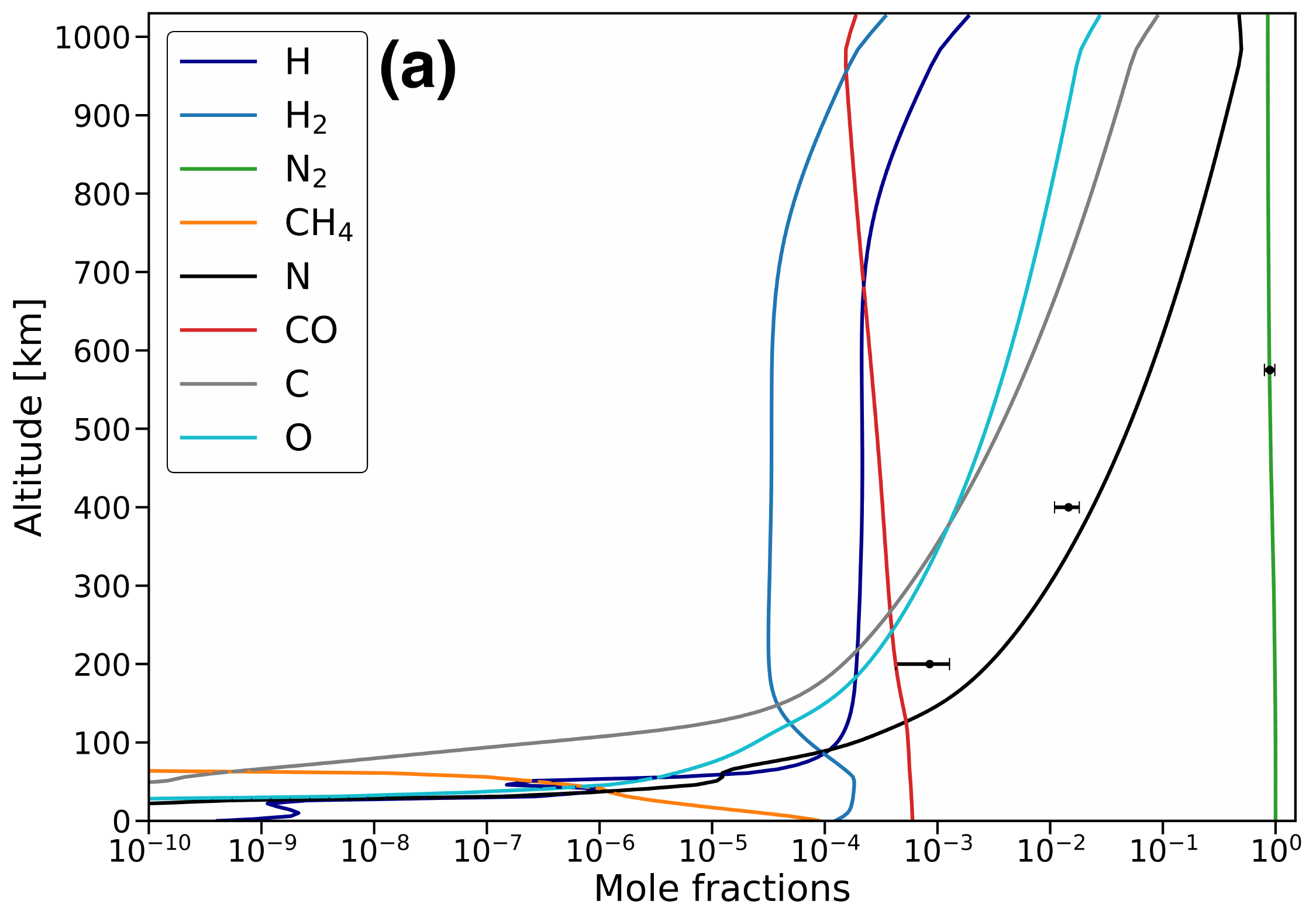}
              \includegraphics[width=0.5\hsize]{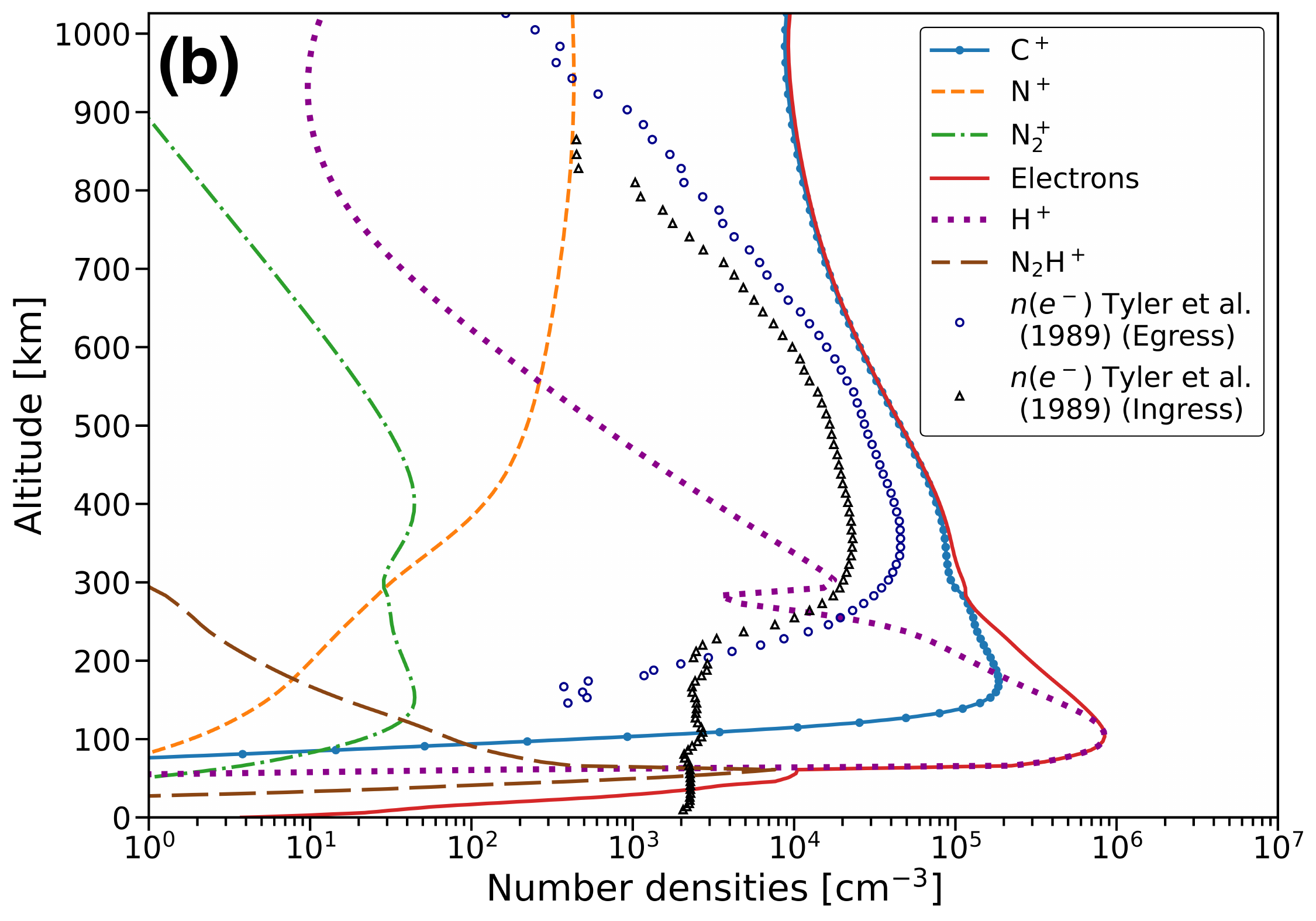}
              }
        \caption{Results from the photochemical model using electron-impact ionization and electron-impact dissociation rates computed by TRANSPlanets with a not renormalized precipitation flux $j^*(E)$. (a) Mole fractions of the main neutral species. The horizontal bars represent Voyager 2 measurements as presented in \citet{krasnopolsky_temperature_1993}. (b) Number density profiles of the main ions. The electron profile can be compared with the profiles presented in \citet{tyler_voyager_1989}.
                }
           \label{fm_fullbeans_neutral+conc_ions_annexe}
  \end{figure*}

  We present results obtained with TRANSPlanets and the photochemical model when the input precipitation flux used in the former model is $j^*(E)$, that is, the flux that is not renormalized. Thus, this flux carries significantly more energy that can be deposited by curvature drift, as computed with Eqs. \eqref{power_deposited} and \eqref{drift_per_pass}, which were taken from \citet{strobel_magnetospheric_1990}. Fig. \ref{fig_prod_elec_fullbeans_annexe} shows the secondary electron production rate, and Fig. \ref{fm_fullbeans_neutral+conc_ions_annexe} plots the mole fractions of the main neutral species and the main ions of the Triton atmosphere. 
  
  Fig. \ref{fm_fullbeans_neutral+conc_ions_annexe} shows that both the mole fraction of atomic nitrogen and the electron density profiles are inconsistent with Voyager 2 data. They were taken from \citet{krasnopolsky_temperature_1993} and \citet{tyler_voyager_1989}.
  
  \section{Outputs from the nominal model using TRANSPlanets}
\label{Appendix_comp_res}

We present in Table \ref{comp_int_col_rates_ME} the integrated reaction rates of the electron-impact ionization and electron-impact dissociation reactions in \citet{benne_photochemical_2022} and in this work.
Table \ref{App_Table_F_uncert} presents the altitude at which the uncertainty on the mole fraction of the main atmospheric species is maximum for this work, along with the uncertainty factor on this mole fraction. 
Finally, Table \ref{comp_int_KCR} compares the key chemical reactions of \citet{benne_photochemical_2022} and this work. 

\begin{table*}[!h]
   \centering          
   \begin{tabular}{@{}cccc@{}}
      \hline
      \begin{tabular}[c]{@{}c@{}}Electron-impact ionization or \\ Electron-impact dissociation reaction\end{tabular} & \begin{tabular}[c]{@{}c@{}}Integrated column\\ rates (cm$^{-2}$.s$^{-1}$)\\ \citet{benne_photochemical_2022}\end{tabular} & \begin{tabular}[c]{@{}c@{}}Integrated column\\ rates (cm$^{-2}$.s$^{-1}$)\\ This work\end{tabular} & \begin{tabular}[c]{@{}c@{}}$\Delta$ \\ relative\end{tabular} \\ \hline
      N$_2$ + e$^-$ $\longrightarrow$ N$_2^+$ + 2e$^-$ & 6.6$\times$10$^7$ & 2.4$\times$10$^7$ & -63.6\% \\
      N$_2$ + e$^-$ $\longrightarrow$ N$^+$ + N($^2$D) + 2e$^-$ & 1.6$\times$10$^7$ & 3.9$\times$10$^6$ & -75.6\% \\
      N$_2$ + e$^-$ $\longrightarrow$ N($^4$S) + N($^2$D) + e$^-$ & 4.9$\times$10$^7$ & 2.1$\times$10$^7$ & -57.1\% \\
      N($^4$S) + e$^-$ $\longrightarrow$ N$^+$ + 2e$^-$ & - & 1.0$\times$10$^5$ & - \\
      N($^2$D) + e$^-$ $\longrightarrow$ N$^+$ + 2e$^-$ & - & 2.3$\times$10$^3$ & - \\
      C + e$^-$ $\longrightarrow$ C$^+$ + 2e$^-$ & - & 9.2$\times$10$^3$ & - \\
      CO + e$^-$ $\longrightarrow$ CO$^+$ + 2e$^-$ & - & 2.0$\times$10$^4$ & - \\
      CO + e$^-$ $\longrightarrow$ C + O($^3$P) + e$^-$ & - & 6.5$\times$10$^3$ & - \\ \hline
      Sum & 1.31$\times$10$^8$ & 4.9$\times$10$^7$ & -62.6\% \\ \hline
      Sum for electron-impact ionization & 8.2$\times$10$^7$ & 2.8$\times$10$^7$ & -65.8\% \\ \hline
   \end{tabular}
   \caption{Comparison of the integrated column rates of the electron-impact ionization and electron-impact dissociation reactions} 
   \label{comp_int_col_rates_ME}
\end{table*}

\begin{table*}[!h]
   \centering          
   \begin{tabular}{@{}lcl@{}}
      \hline
      Species & \begin{tabular}[c]{@{}c@{}}Altitude\\ (km)\end{tabular} & $F(\Bar{y_i})$ \\ \hline
      H$_2$      & 943  & 1.14  \\
      C          & 46   & 2.74  \\
      H          & 31   & 9.17  \\
      CH$_4$     & 153  & 9.28  \\
      C$_2$H$_2$ & 127  & 3.14  \\
      C$_2$H$_4$ & 41   & 10.05 \\
      C$_2$H$_6$ & 86   & 6.98  \\
      HCN        & 438  & 5.39  \\
      O($^3$P)   & 31   & 52.72 \\
      N$_2$      & 1026 & 1.01  \\
      CO         & 1026 & 1.07  \\
      N($^4$S)   & 31   & 26.44 \\
      N($^2$D)   & 115  & 42.01 \\
      H$^+$      & 127  & 5.36  \\
      C$^+$      & 121  & 5.60  \\
      N$^+$      & 923  & 1.59  \\
      N$_2^+$    & 1026 & 1.90  \\
      e$^-$      & 146  & 2.00  \\ \hline
   \end{tabular}
   \caption{Uncertainty factor on the mean abundance $F(\Bar{y_i})$ of the main atmospheric species. \\ These values were computed at the level at which the uncertainty on the abundance of the studied species is maximum. The uncertainty factor $F$ gives the interval $\left [ \frac{\Bar{y}}{F}, \Bar{y}\times F \right ]$ at 1$\sigma$.} 
   \label{App_Table_F_uncert}
\end{table*}

\begin{table*}[!h]
  \begin{adjustwidth}{-1.2cm}{-1.2cm}
   \centering          
   \begin{tabular}{@{}lllll@{}}
      \hline
      \multirow{2}{*}{Reaction} & \multicolumn{2}{c}{This work} & \multicolumn{2}{c}{\citet{benne_photochemical_2022}} \\
       & Species (Production) & Species (Loss) & Species (Production) & Species (Loss) \\ \hline
      CH$_4$+$h\nu$$\longrightarrow$$^1$CH$_2$+H$_2$ & H$_2$(33\%) & CH$_4$(34\%) & H$_2$(34\%) & CH$_4$(35\%) \\
      CH$_4$+$h\nu$$\longrightarrow$CH$_3$+H & H(27\%) & CH$_4$(29\%) & H(30\%) & CH$_4$(30\%) \\
      H+$^3$CH$_2$$\longrightarrow$CH+H$_2$ & H$_2$(47\%) & H(53\%) & H$_2$(48\%) & H(56\%) \\
      CH+CH$_4$$\longrightarrow$C$_2$H$_4$+H & C$_2$H$_4$(73\%); H(21\%) & CH$_4$(23\%) & C$_2$H$_4$(73\%); H(23\%) & CH$_4$(22\%) \\
      H+HCNN$\longrightarrow$$^1$CH$_2$+N$_2$ & N$_2$(23\%) & H(32\%) & N$_2$(17\%) & H(34\%) \\
      N$_2$+$h\nu$$\longrightarrow$N$_2^+$+e$^-$ & e$^-$(48\%); N$_2^+$(55\%) &  & e$^-$(26\%); N$_2^+$(31\%) &  \\
      N$_2$+$h\nu$$\longrightarrow$N($^4$S)+N($^2$D) & N($^4$S)(11\%); N($^2$D)(20\%) &  & &  \\
      N$_2$+e$^-$$\longrightarrow$N$_2^+$+2e$^-$ & e$^-$(39\%); N$_2^+$(45\%) &  & e$^-$(57\%); N$_2^+$(69\%) & \\
      N$_2$+e$^-$$\longrightarrow$N($^4$S)+N($^2$D)+e$^-$ & N($^4$S)(12\%); N($^2$D)(21\%) &  & N($^4$S)(13\%); N($^2$D)(24\%) & \\
      N$_2^+$+H$_2$$\longrightarrow$N$_2$H$^+$+H &  & H$_2$(51\%); N$_2^+$(37\%) &  & H$_2$(52\%); N$_2^+$(15\%) \\
      N$_2^+$+e$^-$$\longrightarrow$N($^4$S)+N($^2$D) & N($^2$D)(16\%) & e$^-$(26\%); N$_2^+$(29\%) & N($^4$S)(11\%); N($^2$D)(20\%) & e$^-$(35\%); N$_2^+$(42\%) \\
      N$_2^+$+e$^-$$\longrightarrow$N($^2$D)+N($^2$D) & N($^2$D)(27\%) & e$^-$(22\%); N$_2^+$(25\%) & N($^2$D)(33\%) & e$^-$(30\%); N$_2^+$(36\%) \\
      N($^4$S)+CN $\longrightarrow$N$_2$+C & C(72\%) & N($^4$S)(20\%) & C(69\%); N$_2$(17\%) & N($^4$S)(37\%) \\
      N($^2$D)+CO$\longrightarrow$N($^4$S)+CO & N($^4$S)(56\%) & N($^2$D)(93\%) & N($^4$S)(48\%) & N($^2$D)(77\%) \\
      H$^+$+HCN$\longrightarrow$HNC$^+$+H &  & H$^+$(22\%); HCN(87\%) &  & H$^+$(28\%); HCN(70\%) \\ \hline
   \end{tabular}
   \caption{Comparison of the key chemical reactions obtained in this work with a slightly modified version of the model of \citet{benne_photochemical_2022}, based on the nominal results. \\ The key chemical reactions are the reactions that contribute more than 10\% of the total integrated production or loss of at least two of the main atmospheric species from this work.} 
   \label{comp_int_KCR} 
  \end{adjustwidth}
\end{table*} 

\newpage

\section{Results from our photochemical model without electron precipitation}
\label{app_C}

\begin{figure}[!h]
   \centering
      {\includegraphics[width=0.65\textwidth]{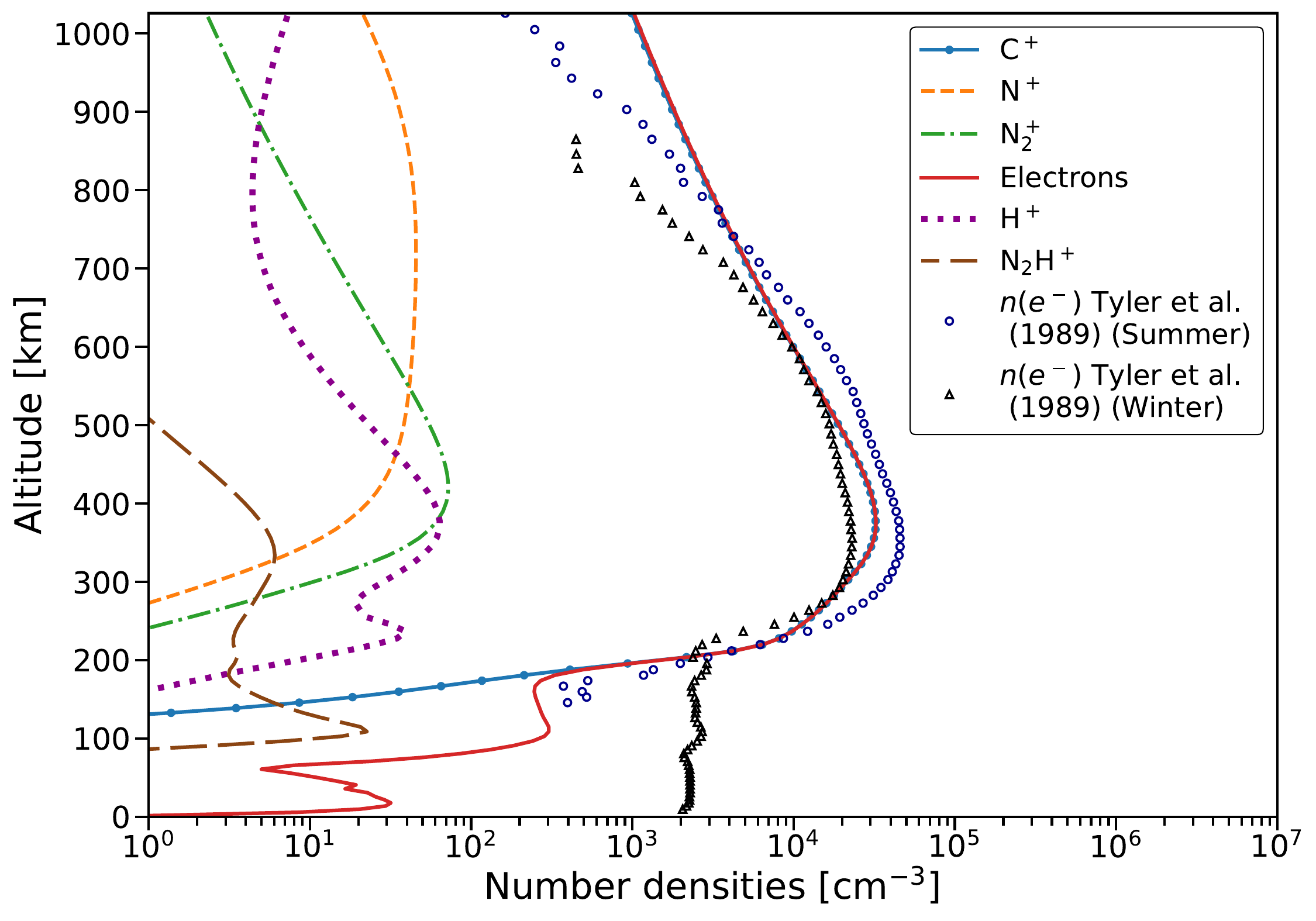}
      }  
   \caption{Number density profiles of the main ions of the Triton atmosphere obtained with the photochemical model without the electron precipitation. The electron profile is compared to Voyager 2 data from \citet{tyler_voyager_1989}. 
    }
   \label{conc_ions_noME}
\end{figure}

In Fig. \ref{conc_ions_noME} we present the number densities of the main ions of the Triton atmosphere obtained with our photochemical model without electron precipitation from the Neptune magnetosphere. Therefore, the rates of the electron-impact ionization and electron-impact dissociation reations are zero. 

\end{appendix}
\newpage 


\bibliographystyle{abbrvnat}
\bibliography{biblio_20221017}
\newpage

\end{document}